\documentclass[useAMS,usenatbib,aas_macros]{mn2e}

\usepackage{lscape}
\usepackage{graphicx}
\usepackage{colortbl}
\usepackage{amssymb}
\usepackage{natbib}
\usepackage{times}
\usepackage{aas_macros}

\newcommand{\Jlw}{$J_{\rm LW}$\ }

\usepackage{color}

\def\simpropto{\lower.2ex\hbox{$\; \buildrel \propto \over \sim \;$}}
\def\ltsim{\lower.5ex\hbox{$\; \buildrel < \over \sim \;$}}
\def\gtsim{\lower.5ex\hbox{$\; \buildrel > \over \sim \;$}}

\begin{document}

\title[Ubiquitous seeding of SMBHs by direct collapse]{Ubiquitous seeding of supermassive black holes by direct collapse}

\author[B. Agarwal, et al.]{Bhaskar Agarwal$^1$\thanks{E-mail:
agarwalb@mpe.mpg.de}, Sadegh Khochfar$^1$, Jarrett L.
Johnson$^1$$^,$$^2$, Eyal Neistein$^1$,
\newauthor Claudio Dalla Vecchia$^1$ and Mario Livio$^3$ \\
$^1$Max-Planck-Institut f{\"u}r extraterrestrische Physik,
Giessenbachstra\ss{}e, 85748 Garching, Germany \\
Theoretical Modeling of Cosmic Structures Group \\
$^2$Los Alamos National Laboratory, Los Alamos, NM 87545, USA \\
Nuclear and Particle Physics, Astrophysics and Cosmology Group (T-2) \\
$^3$Space Telescope Science Institute, 3700 San Martin Drive,
Baltimore, MD 21218, USA}


\date{00 February 2012}
\pagerange{\pageref{firstpage}--\pageref{lastpage}} \pubyear{0000}
\maketitle

\label{firstpage}


\begin{abstract}
We study for the first time the environment of massive black hole (BH) seeds ($\sim 10^{4-5}\ \rm M_{\sun}$) formed via the direct collapse of pristine gas clouds in massive haloes ($\geq 10^{7} \ \rm M_{\sun}$) at $z > 6$. Our model is based on the evolution of dark matter haloes within a cosmological $N$-body simulation, combined with prescriptions for the formation of BH along with both Population III (Pop III) and Population II (Pop II) stars. We calculate the spatially-varying intensity of Lyman Werner (LW) radiation from stars and identify the massive pristine haloes in which it is high enough to shut down molecular hydrogen cooling. 
 In contrast to previous BH seeding models with a spatially constant  LW background, we find that the intensity of LW radiation due to local sources, $J_{\rm local}$, can be up to $\sim10^6$ times the spatially averaged background in the simulated volume and exceeds the critical value, $J_{\rm crit}$, for the complete suppression of molecular cooling, in some cases by 4 orders of magnitude. 
Even after accounting for possible metal pollution in a halo from previous episodes of star formation, we find a steady rise in the formation rate of direct collapse (DC) BHs with decreasing redshift from $10^{-3}$ $\rm Mpc^{-3}$$z^{-1}$ at $z\!=\!12$ to $10^{-2}$ $\rm Mpc^{-3}$$z^{-1}$ at $z\!=\!6$. The onset of Pop II star formation at $z \approx 16$ simultaneously marks the onset of the epoch of DCBH formation, as the increased level of LW radiation from Pop II stars is able to elevate the local levels of the LW intensity to $J_{\rm local} > J_{\rm crit}$ while Pop III stars fail to do so at any time. The number density of DCBHs is sensitive to the number of LW photons and can vary by over an order of magnitude at $z=7$ after accounting for reionisation feedback.
Haloes hosting DCBHs are more clustered than similar massive counterparts that do not host DCBHs, especially at redshifts $z \gtsim 10$. Also, the DCBHs that form at $z>10$ are found to reside in highly clustered regions whereas the DCBHs formed around $z \sim 6$ are more common. We also show that planned surveys with \textit{James Webb Space Telescope} should be able to detect the supermassive stellar precursors of DCBHs.
\end{abstract}


\section{Introduction}
\label{sec:Intro}

It is now an established fact that galaxies host black holes (BH) at their centres \citep{Gebhardt:2000p831, Ferrarese:2000p830, Gultekin:2009p829} with BH masses ranging from $10^{6-9.5}\ \rm M_{\sun}$. The most massive BHs or supermassive black holes (SMBH) are believed to fuel quasars observed as early as $z>6$ \citep[see e.g.][]{Fan:2003p40,Fan:2006p149,Willott:2003p846,Mortlock:2011p447}. This implies that the seeds of these SMBHs must have formed and grown to supermassive scales in the short time before the Universe was even one billion years old. It has also been suggested recently \cite[T11 here after]{Treister:2011p114} that there might be a population of obscured intermediate mass black holes (IMBHs) at $z>6$ \citep[however also see e.g.][ who challenge the claim]{Willott:2011p448, Fiore:2012p834}. However, the origin of these SMBHs or IMBHs in the early Universe is still an open question.

The most obvious way to make the SMBH seeds is from the stellar BHs in the early Universe. Detailed studies have shown that the first generation of stars (Pop III) form from  metal-free gas, comprising mainly of atomic and molecular hydrogen at early times \citep[see reviews by][and references therein]{Bromm:2004p35,Ciardi:2001p84}. Pop III stars with masses in the range $40\ $M$_{\sun}<M_*<140$ M$_{\sun}$ and $M_*>260$ M$_{\sun}$ collapse into a black hole with $M_{\bullet}=0.5-1 \ M_*$ \citep{Heger:2003p23} and accretion of gas onto these stellar BHs offers a natural way to grow SMBHs, given their abundance and early formation times.

This scenario however, has been challenged given that  Pop III remnant BHs may not constantly accrete at or near the Eddington limit, which is likely required for $100 \ \rm M_{\sun}$ seed black holes to reach a mass of $10^9$ M$_{\odot}$ by $z\sim 6$.   Both the radiation from the Pop~III progenitor star \citep[e.g.][]{Yoshida:2006p777,Johnson:2007p48,Alvarez:2009p778} and the radiation emitted in the accretion process itself \citep[e.g.][]{Milosavljevic:2009p779,Park:2011p780,Li:2011p847}, result in feedback which might slow down gas accretion. The constant availability of gas in the halo during the accretion period would also require the haloes to grow rapidly via mergers since episodes of star formation and feedback from supernovae can deplete the gas in such primordial haloes \citep[e.g.][]{Mori:2002p781}. On the other hand, a scenario where the accretion must be super-Eddington for a short period of time has been proposed in order to allow fast BH growth \citep[e.g.][]{Volonteri:2005p793}, which could be a result of the inefficient radiative losses due to the trapping of photons in the accretion disc \citep[see e.g.][]{Begelman:1978p792,Wyithe:2011p835}. 
 
 Another possibility of growing stellar black holes is via mergers of haloes hosting either stars or BHs. A dense cluster or group of stars provides conditions under which frequent mergers can occur, leading to a runaway collapse \citep{Zwart:1999p767} that result in BHs with masses of around $10^5$ M$_{\odot}$.  Mergers of Pop III seed BHs at high redshifts can also, in principle, build up supermassive BHs \citep{Tanaka:2009p838}, although slingshot effects and merger time scales pose problems for this scenario \citep[see e.g. the reviews by][and references therein]{Natarajan:2011p90, Volonteri:2010p30}.  
 
An alternative scenario is to make seed BHs with an initial $M_{\bullet}=10^{4-5}$ M$_{\sun}$ via the direct collapse of pristine gas in haloes with $T_{\rm vir} \ge 10^4$ K \citep[see e.g.][]{Eisenstein:1995p870,Oh:2002p836,Bromm:2003p22,Koushiappas:2004p871,Lodato:2006p375}. The key idea is to keep the haloes free of molecular hydrogen so that the gas collapses isothermally only via atomic hydrogen. For the gas collapse to proceed without fragmenting into stars, it also has to redistribute its angular momentum and various processes have been suggested in order to allow this, as explained below.

In low spin haloes the gas settles down into a disc where the angular momentum can then be redistributed via gravitational instabilities, hence keeping the Toomre parameter close to unity and preventing the disc from further fragmentation \citep[LN06 hereafter]{Lodato:2006p375}. The central core of $M=10^{4-5} \ \rm M_{\sun}$, fed by the streams resulting from the non-axisymmetric disc instabilities, ultimately collapses into a BH with a similar mass. An important feature of LN06 is that they explicitly link the dark matter halo properties, like spin and virial temperature, to the properties of the BH seed. Their model predicts the required ratio of the gas temperature to the virial temperature and the maximum halo spin which determines the final mass of the BH seed.

The redistribution of angular momentum can occur via the `bars-within-bars' scenario as explored by \cite{Begelman:2006p75} where the gas collapses into a dense self-gravitating core surrounded by an envelope supported by radiation pressure. The gas finally cools and collapses catastrophically via neutrino emission into a central BH with an intermediate stage of a quasi-star \citep{Begelman:2008p672}. 

\cite{Spaans:2006p58} showed that if the collapse of gas (comprised of atomic H) in such haloes proceeds via an equation of state with a polytropic index larger than unity, Lyman-alpha photons can get trapped in highly dense regions owing to the large optical depth of the medium. The time required for the Lyman-alpha photons to escape the medium becomes larger than the free fall time of the gas which prevents the gas from cooling and forming Pop III stars. Hence, the collapse can result in a massive BH which is of the order of $3-20 \%$ of the total baryonic mass of such haloes.

Also, \cite{Regan:2009p776} explored the gas collapse in rare atomic cooling haloes which could in principle host a DCBH in cosmological hydrodynamic simulation. They find cases where the inflow rates are high enough ($>1 \ \rm M_{\sun} yr^{-1}$) to allow for the formation of massive BH seeds.

All these scenarios end in a \textit{direct collapse} black hole (DCBH) with $M_{\bullet}\sim10^{4-6}$ M$_{\sun}$. Another alternative scenario includes the formation of a supermassive star (SMS) in an intermediate step on the way to the formation of a DCBH \citep{Begelman:2010p872}. For this to occur the gas does not only need to be free of $\rm H_2$ and metals but the accretion rate onto the SMS needs to be high enough to allow the rapid growth to $10^{4-6}$ M$_{\odot}$ \citep{Begelman:2010p872, Johnson:2012p874}.

Although these scenarios take place in haloes with $\rm T_{\rm vir}>10^4\ \rm K$, which are mostly composed of atomic hydrogen, molecular hydrogen can form in these haloes when the densities are high enough to allow three-body hydrogen interactions. Such high particle densities are found at the halo centre and during the end stages of gas collapse. Hence these scenarios require a critical level of H$_{\rm 2}$ photo-dissociating Lyman Werner (LW) radiation ($h\nu = 11.2-13.6$~\ \rm eV) in order to keep the abundance of H$_{\rm 2}$ molecules very low, as otherwise H$_{\rm 2}$ cooling will lower temperatures to $\rm T\approx 200\ K$, thereby reducing the Jeans mass and leading to fragmentation of the gas cloud, which would ultimately result in star formation instead of a central BH seed.

The main challenge in all the above DCBH formation scenarios is to reach the critical level of Lyman Werner radiation required to dissociate H$_{\rm 2}$ molecules in the halo. Typical levels of a smooth uniform LW background, $ J_{\rm bg}$, range from $0.001-0.1$ (where $J$ is expressed in units of 10$^{-21}$ $\rm erg\ \rm s^{-1}\rm cm^{-2}\rm Hz^{-1}\rm sr^{-1}$) and depends on the stellar density at a given redshift \citep{Greif:2006p99}, whereas the critical value, $J_{\rm crit}$, required for direct collapse is $\sim 30$ (from Pop II) and $\sim 1000$ (from Pop III) \citep[][ CS10 and WG11 herafter]{Shang:2010p33,WolcottGreen:2011p121}. It has been argued that a halo can be exposed to a radiation level higher than $J_{\rm crit}$ if it lives close to a star forming region \citep[D08 hereafter]{Dijkstra:2008p45}. They use an analytical approach employing Poisson statistics and extended Press-Schetcher mass functions to model their halo distribution which accounts for clustering of the DM haloes and the spatial distribution of LW sources. 

Previous studies of DCBH formation have either assumed a spatially constant LW background \citep{Regan:2009p776,Petri:2012p1335} or a spatially varying LW background using analytical prescriptions for clustering of sources D08. The latter study showed that the clustering of sources plays a crucial role in elevating the levels of LW radiation above the critical value required for DCBH formation. While it is important to model the clustering of sources properly, it is also crucial to know whether a halo, which is exposed to the critical level of LW radiation, had previous episodes of star formation which enriched the gas in the halo with metals. In contrast to D08, \cite{Petri:2012p1335} attempted to model the merging histories of haloes using Monte-Carlo merger trees however, they did not account for the self consistent build up of the spatially varying LW radiation field. 

Due to the importance of LW feedback at high redshifts, some recent studies have explored the effects of LW radiation on early structure formation \citep[e.g.][]{Kuhlen:2011p787}, Pop III star formation \citep{SafranekShrader:2012p927,Whalen:2008p785,Ricotti:2008p782}, the evolution of pair instability supernovae \citep{Wise:2012p458,Hummel:2011p456} and also on the formation of SMBH seeds by Pop III stars \citep{Devecchi:2012p1015}.

In this paper we simultaneously follow the build up of the spatially varying LW radiation field as well as track the enrichment histories of dark matter (DM) haloes in a cosmological DM only, $N$-body simulation using a semi-analytical model (SAM). We investigate the conditions under which the LW intensity seen by an individual halo will reach a value $\gtsim J_{\rm crit}$ and we describe the resulting consequences for the formation of seed BHs via direct collapse. The aim of our work is to determine the plausibility of the existence of DCBH sites and probe the clustering features of such haloes.

This paper is organised as follows. We describe the simulation and our model in the next section (Sec. \ref{sec:Method}) followed by which the results of our work are presented in Sec. \ref{sec:Results}. The observability of the supermassive stellar seeds of DCBHs by the \textit{James Webb Space Telescope} (JWST) is discussed in Sec. \ref{sec:Observations}. Finally we present the summary and discussion of our work in Sec. \ref{sec:Discussion}.

\section{Methodology}
\label{sec:Method}

In the following section(s) we describe our SAM that models the build up of the LW radiation field on top of our $N$-body simulation. We model both Pop III and Pop II star formation and include a prescription for the evolution of the star forming and non-star-forming gas within an individual halo. This allows us to track the star formation histories of the haloes and account for the LW photon travel times which is needed in order to self consistently model the global and spatial level of the LW radiation at each point in our box.

\subsection{The $N$-body simulation}  

We use a DM only $N$-body simulation with $768^3$ particles in a $3.4$ Mpc$h^{-1}$ co-moving periodic box using the \textsc{gadget} code \citep{Springel:2001p10, Springel:2005p667}. We assume a $\Lambda$CDM cosmology with $\Omega_0=0.265$, $\Omega_b=0.044$, $\Omega_\Lambda=0.735$, $h=0.71$ and $\sigma_8=0.801$ consistent with the WMAP 7 results \citep{Komatsu:2011p409}. The resulting individual DM particle mass is $\rm 6500\ M_{\sun}$$h^{-1}$. Merger trees are constructed on the \textsc{subfind} output \citep{Springel:2001p10} using the same method as in \cite{Springel:2005p851}. Information on each subhalo includes its mass as assigned by \textsc{subfind}, along with its host friends-of-friends (FoF) mass. The smallest resolved DM halo contains at least 20 particles, which corresponds to $ 1.3\times 10^5 \ \rm M_{\sun}$$h^{-1}$. We run the simulation down to $z=6$ and the snapshots are taken a few tens of Myr apart.
As in \cite{Springel:2005p851} and \cite{Croton:2006p852}, our merger trees are based on subhaloes. Note that we shall use the term \textit{halo} instead of \textit{subhalo} in the remainder of this work for the sake of simplicity.

At a given snapshot, we label the haloes as \textit{minihaloes} and \textit{massive haloes} if their virial temperature is $2000 K\leq T_{\rm vir} < 10^4\ \rm K$ and $T_{\rm vir} \geq 10^4\ \rm K$, respectively. Also, $J$ or the combination of the variable with any superscript/subscript explicitly implies \Jlw in units of $10^{-21}$~erg$^{-1}$s$^{-1}$cm$^{-2}$Hz$^{-1}$sr$^{-1}$ unless specified otherwise.

We define the infall mass, $M_{\rm infall}$, of the halo as its mass at the last snapshot where it was the most massive subhalo within its FoF halo. We did this by tracking the halo's main progenitor branch back in time. The infall redshift is defined as the redshift when the infall mass was found.

We use the relations from \cite{Barkana:2001p60} for the virial temperature, virial radius, $R_{\rm vir}$, and circular velocity, $V_{\rm c}$, of a halo
\begin{eqnarray}
\label{eq.Tvir}
 T_{\rm vir} = 1.98 \times 10^4\left(\frac{\mu}{0.6}\right) \left(\frac{M_{\rm infall}}{10^8h^{-1}\mbox{M}_{\sun}}\right)^{\frac{2}{3}} \times \nonumber \\  
 \left(\frac{\Omega_0}{\Omega_m(z)}\frac{\Delta_c}{18\pi^2}\right)^{\frac{1}{3}} \left(\frac{1+z}{10}\right) \mbox{K} \ ,
\end{eqnarray}
\begin{eqnarray}
\label{eq.Rvir}
 R_{\rm vir}=0.784\left(\frac{M_{\rm infall}}{10^8 h^{-1}\mbox{M}_{\sun}}\right)^{\frac{1}{3}}\left(\frac{\Omega_0}{\Omega_m(z)}\frac{\Delta_c}{18\pi^2}\right)^{-\frac{1}{3}} \times\nonumber \\\left(\frac{1+z}{10}\right)^{-1} \mbox{kpc} \ , 
\end{eqnarray}
\begin{equation}
\label{eq.Vc}
V_{\rm c}=\left(\frac{GM_{\rm infall}}{R_{\rm vir}}\right)^{1/2} \ ,
\label{eq.Vcir}
\end{equation}
where $\mu$ is the mean molecular weight ($1.22$ for neutral primordial gas), $\Omega_0$ is the matter density of the Universe at $z=0$, $\Omega_m(z)$ is the matter density of the Universe as a function of redshift and $\Delta_c$ is the collapse over-density and $z$ denotes the infall redshift as computed from our trees.

\subsection{Star formation}
\label{sec:starformation}
In order for the first star to form out of the gas in a virialised pristine halo, the cooling time, $t_{\rm cool}$, for the gas must be less than the Hubble time, $t_{\rm Hubble}$. The primordial gas mostly comprises of either atomic or molecular hydrogen and the cooling time depends on their respective cooling functions. Atomic hydrogen cooling is effective at $T>10^4\ \rm K$ whereas molecular cooling can operate at lower temperatures. In our model, since we probe the universe at $z\le30$, we use the results from the study by \cite{Tegmark:1997p937} which showed that the critical fraction of $\rm H_2$ molecules required in order to satisfy the condition $t_{\rm cool}<t_{\rm Hubble}$ is found in haloes with $T_{\rm vir} \sim 2000\ \rm K$ at $z=25$. Hence, the first star to form from a pristine gas cloud would be a Pop III star forming in a minihalo. The metals ejected from the first Pop III star would be enough to pollute the gas and Pop II stars could form subsequently in the same halo \citep[e.g.][]{Maio:2010p15}. We discuss the Pop III and Pop II star formation in more detail in the following sections.

As explained above, since it is critical to resolve minihaloes of mass $\sim 10^{5-7} \ \rm M_{\sun}$, this requirement limits the volume that we can probe with sufficient resolution in our study. We plot the mass functions of the FoF and subhaloes in our work at $z=6$ in the Appendix. 

\subsubsection{Pop III stars}

In our model, we allow a single episode of instantaneous Pop III star formation in pristine haloes with with $T_{\rm vir} \ge 2000\ \rm K$ \citep{Tegmark:1997p937,Maio:2010p15}. Here, we consider a halo to be pristine if none of its progenitors have hosted a star in the past. In addition, our implementation of LW feedback, as explained in Sec. \ref{sec:threshmass}, regulates which pristine haloes form Pop III stars. Hence the non-Pop III-forming minihaloes can later to grow into pristine massive haloes.

We assume a Salpeter IMF \citep{Salpeter:1955p861} with a mass range between 100 and 500 $\rm M_{\sun}$ and assume that one Pop III star forms per minihalo \citep[see e.g.][]{Bromm:2004p35}. However, in massive pristine haloes ($T_{\rm vir} \gtsim 10^{4}$ K) we form 10 stars following a Salpeter IMF, with mass cut offs at 10 and 100 $\rm M_{\sun}$ \citep[e.g.][]{Greif:2008p377,Johnson:2008p29,Wise:2008p789}. Our choice of IMFs and mass cut-offs in both minihaloes and massive haloes is primarily to maximise the LW output from the stars. Forming multiple lower mass stars as opposed to a single very massive star gives an upper limit to the amount of LW radiation that can be emitted from a massive pristine halo as, for instance, the number of LW photons produced by five 100 M$_{\odot}$ stars is larger than for one 500 M$_{\odot}$ star (see Sec. \ref{sec.JLW}). 

Since the formation time for a Pop III star is few Myr \citep[e.g.][]{Bromm:2009p32} and our snapshots are $\approx 10\ \rm Myr$ apart, Pop III stars are assigned a time of birth and distributed uniformly in the time interval between two subsequent snapshots (see Appendix). The masses of individual stars within a pristine halo are generated randomly following the respective IMFs assumed.

\subsubsection{Pop II stars}
\label{sec:popii}
The second generation of stars, Pop II, is also expected to exist at high redshifts within metal-enriched regions \citep[e.g.][]{Wise:2008p789,Greif:2010p57}. These stars are metal rich as compared to Pop III but have metallicities much smaller than the solar metallacity, $\rm Z_{\sun}$. The metals ejected from Pop III stars pollute the host and neighbouring haloes via stellar and SN winds \citep{Mori:2002p781,Maio:2011p104}. Any further collapse of the polluted gas in the haloes would result in cooling to lower temperatures, thereby reducing the Jeans mass and forming metal-enriched stars with lower masses than the Pop III stars \citep[e.g.][]{Clark:2008p843,Smith:2009p795}. The critical metallicity at which the transition occurs from Pop III to Pop II ranges from $10^{-4}$ to $10^{-6}\ \rm Z_{\sun}$ \citep[e.g.][]{Frebel:2007p826}. For simplicity, we consider a halo that has hosted a Pop III star (or merged with a halo hosting or having hosted a Pop III or Pop II star) polluted with metals and a possible site for Pop II star formation \cite[see e.g.][]{Johnson:2010p18}.

Since metals are the coolants required for making Pop II stars, we assume that a large enough potential well would be required to constrain the metals ejected from Pop III SNe and additionally add a constrain by setting the threshold halo mass\footnote{See Section \ref{sec:Discussion} for a critical view on how the choice of the Pop II threshold mass sets the clock for DCBH formation.}
 for Pop II star formation to $10^8 \ \rm M_{\sun}$ \citep[e.g.][]{Kitayama:2004p669,Whalen:2008p785}. Within these candidate haloes, we assume that the baryons can exist in either of the three phases i.e. non-star-forming gas, star-forming gas or stars. Below we describe the transition between these phases which ultimately regulates the Pop II star formation in a halo.

\begin{itemize}

\item \textit{Non-star-forming gas phase}: 
We assume in our model that once a DM halo crosses our resolution limit of 20 particles, it is initially comprised of non-star-forming gas, $M_{\rm hot} = f_{\rm b} M_{\rm DM}$, where $f_{\rm b}=0.16$ is the universal baryon fraction and $M_{\rm DM}$ is the halo's current DM mass.

While the DM halo grows between two snapshots, we add non-star-forming gas to the halo by calculating the accretion rate, $\dot M_{\rm acc}$, defined as 
\begin{equation}
\dot M_{\rm acc} \equiv  \frac {f_{\rm b} \Delta M_{\rm DM} - M_{\rm *,p} - M_{\rm out,p}}{\Delta t} \ ,
\label{eq.acc}
\end{equation}
where $\Delta M_{\rm DM}$ is amount by which the DM halo grows between two snapshots which are $\Delta t$ apart, $M_{\rm *,p}$ and $M_{\rm out,p}$ represent the total stellar mass and net mass lost in previous SN outflows summed over the incoming merging haloes respectively.
\\
\item \textit{Star forming gas phase}:
In order for the gas to form stars, it must cool and collapse within the halo. We model the transition from the non-star-forming gas phase to the star-forming gas phase, $M_{\rm cold}$, by allowing $M_{\rm hot}$ to collpase over the dynamical time of the halo, $t_{\rm dyn}=\frac{R_{\rm vir}}{V_c}$. This estimate is justified by the fact that at such high redshifts, the radiative-cooling time is shorter than the dynamical time of the halo.
\\
\item \textit{Star formation law}:
We then model the Pop II star formation via a Kennicutt-type relation \citep{Kennicutt:1998p143} 
\begin{equation}
\dot{M}_{*,\rm II}=\frac{\alpha}{0.1 t_{\rm dyn}}M_{\rm cold} \ ,
\label{eq.td0.1}
\end{equation}
where $\alpha$ is the star formation efficiency (SFE). The factor $0.1t_{\rm dyn}$, which is the star formation time scale, is motivated by \cite{Kauffmann:1999p839, Mo:1998p837} . 

Local observations indicate an $\alpha \sim 0.2$, however, at this stage it is not clear if this also holds at high redshifts ($z>6$), \citep{Khochfar:2011p840}. We therefore treat $\alpha$ as a free parameter and normalise our model to the observations of the cosmic SFRD at $z\gtsim6$.
\\
\item \textit{Outflows}:
In addition to star formation, we also consider the SN feedback processes in a star forming halo. We model the outflow rate of gas from a Pop II star forming halo via the relation
\begin{equation}
\dot{M}_{\rm out} =\gamma \  \dot{M}_{*,\rm II} \ ,
\label{eq.outflowmain}
\end{equation}
where
\begin{equation}
\gamma = \left(\frac{V_{\rm c}}{V_{\rm out}}\right)^{-\beta} \ .
\label{eq.outflow}
\end{equation}
The functional form of $\gamma$ is taken from \cite{Cole:2000p863}. We normalise the parameters in Eq. \ref{eq.outflow} to the results of the high resolution hydrodynamical simulations of the high redshift Universe (Dalla Vecchia and Khochfar 2012, in prep) and for the halo mass range considered in this work. This yields an outflow velocity $V_{\rm out}=110 \ \rm km\, s^{-1}$ and $\beta = -1.74$ resulting in typical values of $\gamma \approx 20$.
We assume that the outflows are generated in the star-forming gas phase and hence $M_{\rm out}$ is subtracted from $M_{\rm cold}$.
\\
\item \textit{Implementation}:
Each time interval between two consecutive snapshots, $\Delta t$, is divided into 100 smaller intervals and the following set of coupled differential equations (along with Eqs. \ref{eq.acc}, \ref{eq.td0.1} and \ref{eq.outflowmain}) for the individual baryonic components are numerically solved over the small time steps: 
\begin{equation}
\dot{M}_{\rm cold} = \frac{M_{\rm hot}}{t_{\rm dyn}} - \dot{M}_{*,\rm II} - \dot{M}_{\rm out} \ ,
\end{equation}
\begin{equation}
\dot{M}_{\rm hot} = -\frac{M_{\rm hot}}{t_{\rm dyn}} +  \dot M_{\rm acc} \ .
\end{equation}

\end{itemize}

\subsection{Impact of LW radiation on star formation and direct collapse}
\label{sec:threshmass}

\begin{figure}
\includegraphics[width=0.47\textwidth,height=0.34\textwidth,trim=15mm 0 0 0]{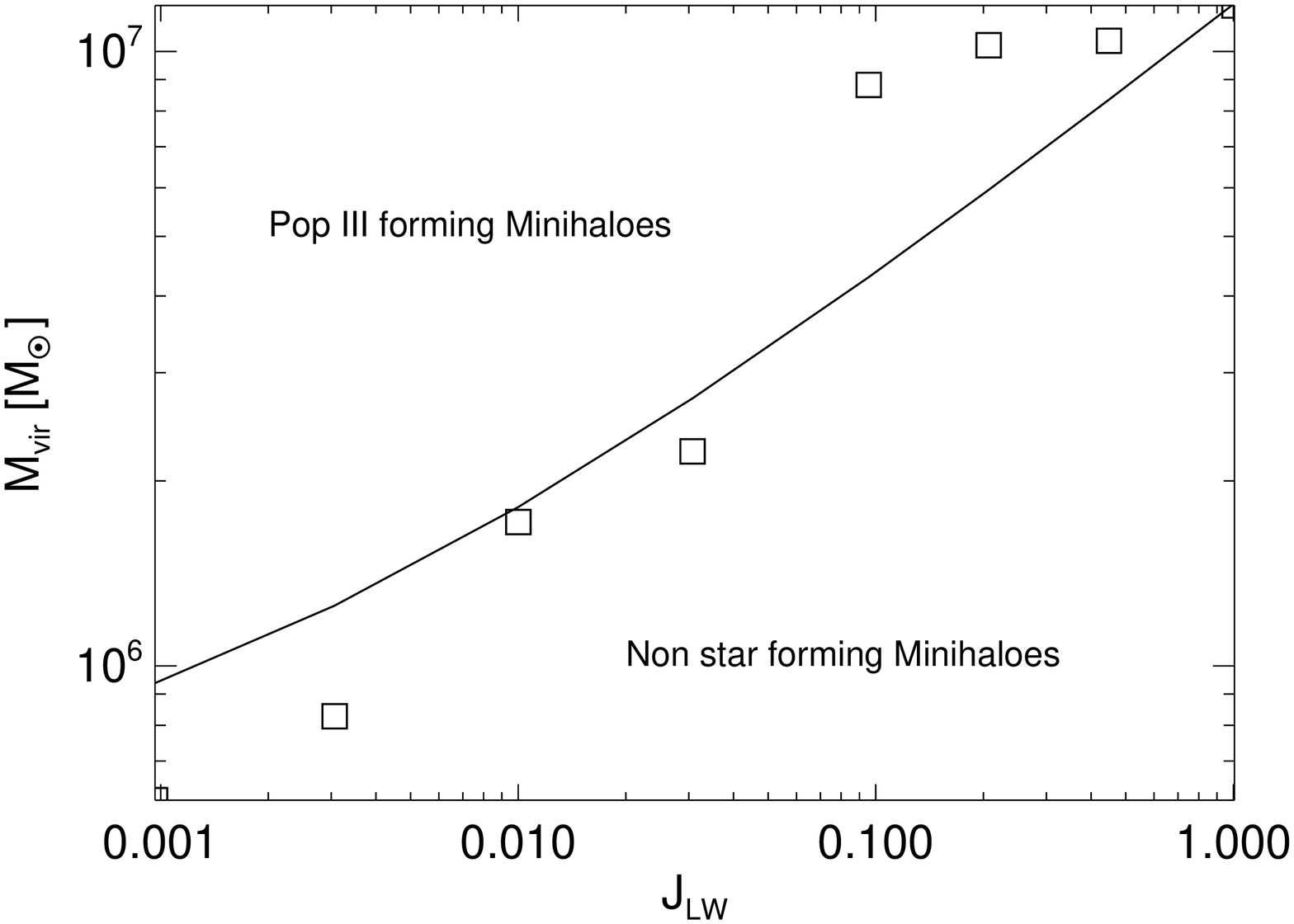}
\caption{The fit used in our model based on the \protect\cite{OShea:2008p41}, Fig. 3(c) (square-symbols) which determines the minimum mass of a pristine minihalo that can host a Pop III star for a given level of external LW radiation it is exposed to. For a metal-free minihalo exposed to a given $J_{\rm LW}$, if its mass lies above the line it is considered Pop III star forming. The solid line represents Eq. \ref{eq.mJvirial}.}
\label{osheanorman}
\end{figure}
Once the first generation of stars form in the Universe, the effects of LW radiation become important for subsequent star formation \citep[e.g.][]{Haiman:2000p87,Omukai:2001p128}. Even a small, uniform $J_{\rm LW}\approx 0.01$ from these stars can affect  Pop III star forming minihaloes by dissociating a fraction of the H$_{\rm 2}$ molecules and preventing the gas from cooling and collapsing \citep{Machacek:2001p150,Yoshida:2003p51,Wise:2007p89, OShea:2008p41}. The amount of H$_{\rm 2}$ molecules that can be dissociated depends directly on the LW background it is exposed to.

The minimum mass, $M_{\rm crit}$, of a pristine halo in which the gas is able to cool, collapse and form Pop III stars in the presence of a given external LW radiation intensity can be approximated by
 \begin{eqnarray}
M_{\rm crit} = {\psi} \left(1.25 \times 10^5 + 8.7 \times 10^5 \left(\frac{ J_{\rm LW} } { 4\pi} \right)^{0.47}\right) \ ,
\label{eq.mJvirial}
\end{eqnarray}
where the expression within the brackets is the functional fit to the numerical simulations carried out by \cite{Machacek:2001p150}. The correction factor $\psi$ has been set to 4 following the higher resolution simulations of \cite{OShea:2008p41}, as shown in Figure \ref{osheanorman}.

One might argue from  Fig. \ref{osheanorman} that $J_{\rm LW}=1$ is sufficient to set the threshold mass to $10^7 \ \rm M_{\sun}$, which is the mass beyond which pristine haloes can cool via atomic hydrogen and hence, direct collapse should ensue. However, detailed simulations by CS10 and WG11 show that only the H$_{\rm 2}$ molecules in the outer regions of such halos are dissociated and a considerable fraction of molecular hydrogen ($\sim 10^{-3}$) still exists in the central region of the halos in the presence of such low levels of $J_{\rm LW}$. 
In order to prevent star formation in the central parsec region of the halo, it is essential to bring down the $\rm H_2$ fraction in the gas to $10^{-8}$ which can be achieved by a $J^{\rm III}_{\rm crit} = 1000$ from Pop III stars (WG11) or  $J^{\rm II}_{\rm crit}=30-300$ from Pop II stars (CS10). The difference in the values of $J_{\rm crit}$ for Pop III and Pop II stars is due to the difference in the spectral shapes of the two stellar populations. As shown by CS10, the lower value of $J^{\rm II}_{\rm crit}$ can be partly attributed to the fact that the $\rm H^-$ dissociation rate from Pop II stars is $\approx 10,000$ times larger than that from Pop III stars, due to the softer shape of the Pop II spectrum at $0.76\ \rm eV$. Since, H$^-$ is a precursor to H$_{\rm 2}$ formation, destruction of H$^-$ is critical as it results in a lower rate of H$_{\rm 2}$ production.\footnote{
H$^-$ is is dissociated by the following photoreaction:
\begin{equation}
H^- + h\nu \rightarrow H + e^- 
\end{equation}
The dissociation rate  can be written as $k_{28}=10^{-10}$  s$^{-1}$ $\alpha \  J_{\rm LW}$. Here, $\alpha_{\rm III} =0.1$ for Pop III stars and $\alpha_{\rm II}=2000$ for Pop II stars. Since $\rm H^-$ can lead to $\rm H_2$ formation, this reaction is of prime importance in order to keep the gas at a low $\rm H_2$ fraction.}

Thus, if a metal free halo with $T_{\rm vir} > 10^4\ \rm K$ is exposed to a critical level of LW radiation, a direct collapse can ensue. In this scenario, the cooling is suppressed and the gas stays at $\approx 8000K$. Due to the large Jeans mass and high accretion rates that these high temperatures imply, a SMS forms and subsequently collapses into a BH \citep[e.g.][]{Bromm:2003p22,Johnson:2012p874}. The central BH then continues to accrete, embedded in an envelope of gas, at super-Eddington efficiencies reaching a quasi-star state and collapsing into a $M_{\bullet}=10^{4-5} \ \rm M_{\sun}$ black hole \citep{Begelman:2008p672}. 

To summarize, only metal-free minihaloes with masses larger than $M_{\rm crit}$ (see Eq. \ref{eq.mJvirial}) are considered to be Pop III star-forming. Also, if these minihaloes are exposed to a $J_{\rm LW} \geq 1$ then they are considered to be non Pop III star-forming. The metal-free massive haloes still make Pop III stars irrespective of $J_{\rm LW}$, given that $J_{\rm LW} < J_{\rm crit}$ otherwise they can be considered to be DC candidates. Also, the Pop II haloes are unaffected by any value of $J_{\rm LW}$ since they are polluted by previous generations of stars, have $M_{\rm infall}>10^8 \ \rm M_{\sun}$ and the coolants are metals. Therefore, the Pop~II star-forming criteria for these haloes is that $M_{\rm infall} > 10^8 \ \rm M_{\sun}$ and that the halo has hosted stars previously or has undergone a merger with a previously star-forming halo. 

Hence, for pristine minihaloes, i.e. haloes with masses in the range corresponding to $2000 \leq T_{\rm vir} < 10^4\ \rm K$,
\[  \left. \begin{array}{l l}
	M \ge M_{\rm crit}\\
         J_{\rm LW} < 1 \end{array} \right\} \rm Pop \ III \] 
The other pristine minihaloes that do not satisfy the above conditions can not form Pop III stars.

For pristine massive haloes i.e. haloes with masses corresponding to $T_{\rm vir} \geq 10^4\ \rm K$,
\[ \left. \begin{array}{l}
         J_{\rm LW} < J_{\rm crit} \end{array} \right\} \rm Pop \ III  \] 
\[  \left. \begin{array}{l}
         J_{\rm LW} \ge J_{\rm crit} \end{array} \right\} \rm DCBH \] 
%
\begin{figure}
\includegraphics[width=0.47\textwidth,height=0.34\textwidth,trim=15mm 0 0 0]{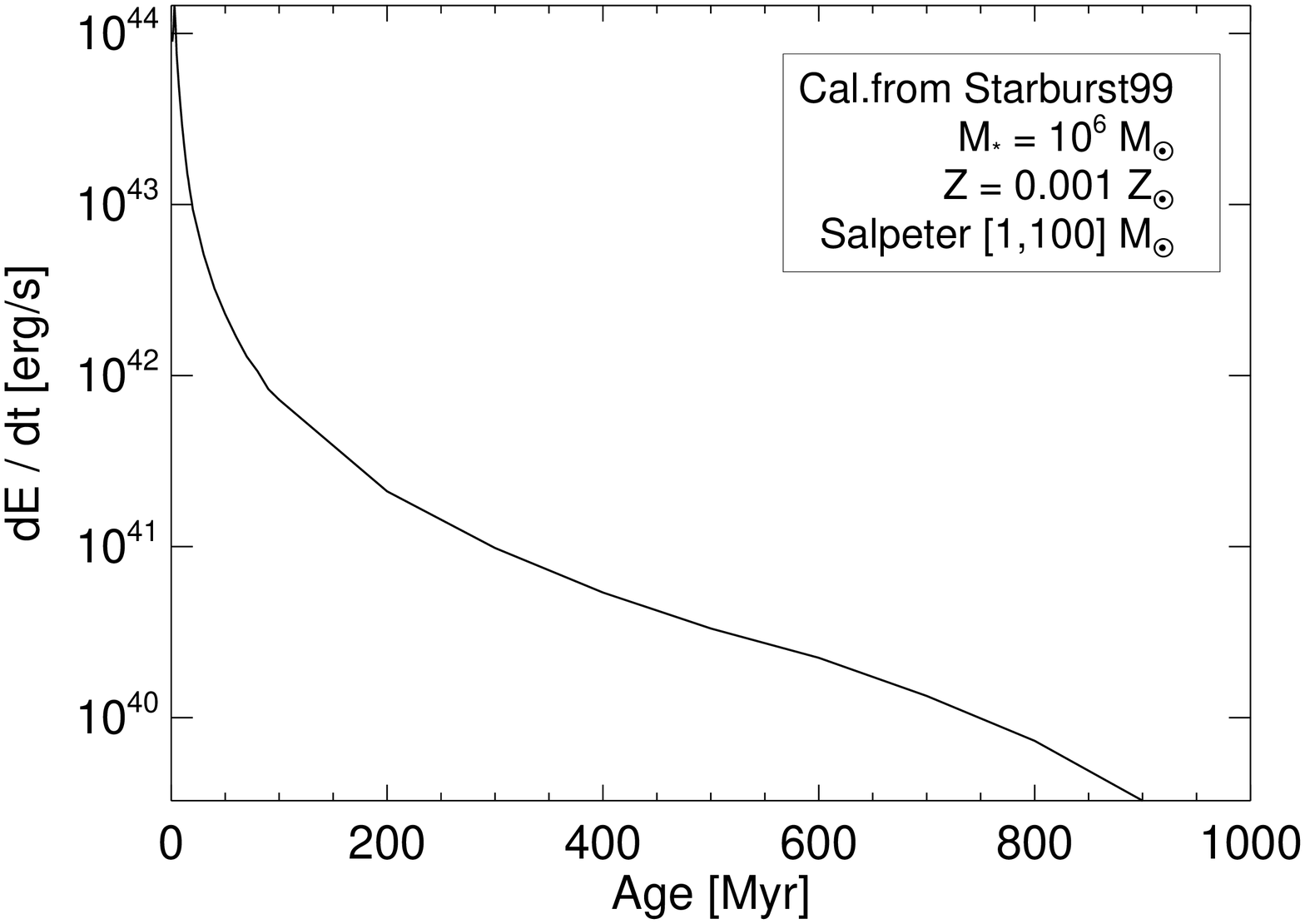}
\caption{Emission in the LW band from a Pop II population as a function of its age computed using the data from \textsc{starburst99} catalogue \protect{\citep[][Fig. 7e]{Leitherer:1999p112}}.}
\label{fig.ST99}
\end{figure}
\subsection{\Jlw calculation}
\label{sec.JLW}

We describe our calculation of the mean and local LW intensities from both the Pop III and Pop II stellar populations in this section. 

\subsubsection{Mean \Jlw  calculation}
\label{sec:MeanJ}

The first stellar populations in the Universe mark the onset of the ultra-violet (UV) background which has a negative effect on star formation as described in the previous sections. The LW photon horizon is larger than our box size \citep[$\sim$10 Mpc,][]{Haiman:2000p87}. Therefore the contribution to the background must come also from outside our simulated volume. In order to account for this, we assume that the SFRD in our volume is representative of a larger cosmological volume.
The mean LW background in our volume, is then assumed to exist everywhere in the Universe and is assumed to be the minimum level of LW radiation that a halo is exposed to at any given redshift.
It can be computed following the formulae in \cite{Greif:2006p99}:
\begin{eqnarray}
\label{eq.JLWIII}
J^{\rm III}_{\rm bg}&\simeq &f_{\rm esc} \frac{hc}{4\pi m_{\rm H}}\eta^{\rm III}_{\rm LW}\rho^{\rm III}_*(1+z)^3 \ ,\\
\label{eq.JLWII}
J^{\rm II}_{\rm bg}&\simeq &f_{\rm esc} \frac{hc}{4\pi m_{\rm H}}\eta^{\rm II}_{\rm LW}\rho^{\rm II}_*(1+z)^3 \ ,
\end{eqnarray}
where $f_{\rm esc}$ is the escape fraction of LW photons from the halo, $\rho^{\rm III}_*,\rho^{\rm II}_*$ denote the comoving density of Pop III and Pop II stars respectively at the given redshift $z$, $\eta_{\rm LW}$ is the number of LW photons per stellar baryon \citep[$\eta^{\rm III}_{\rm LW}=10^4$ and $\eta^{\rm II}_{\rm LW}=4\times10^3$ for the assumed IMFs in our study and as in][]{Greif:2006p99} and $h$, $c$, $m_{\rm H}$ are the Planck's constant, speed of light and mass of a hydrogen atom respectively. In our model, the parameters $\rho^{\rm III}_*,\rho^{\rm II}_*$ are computed by checking if a star or stellar population is active at the current snapshot. Each stellar source in our model is given a time of birth, which is the epoch at which the star is formed and a lifetime depending upon the mass (see Appendix for more details).
\begin{table}
\caption{Functional fits used to compute the radiation output and age of Pop III stars. Values are taken from \protect\cite{Schaerer:2002p21} for a zero metallicity, no mass loss case.}
\begin{center}
\begin{tabular*}{0.45\textwidth}{@{\extracolsep{\fill}}lllll}
\hline
$\mathbf{X}$ & $\mathbf{a_0}$ & $\mathbf{a_1}$ & $\mathbf{a_2}$ & $\mathbf{ a_3}$ \\ 
\hline
$\mathbf{\rm age \ (Myr)}$ & 9.785 & -3.759 & 1.413 & -0.186 \\ 
$\mathbf{Q_{H_2}}$ & 44.03 & 4.59 & -0.77 & - \\ 
$\mathbf{Q_{\rm H}}$ &43.61 &4.90 & -0.89 & - \\ 
\hline
\end{tabular*}
\end{center}
\label{table.func fits}
\end{table}
%
\subsubsection{Spatial variation of \Jlw}
\label{sec:spatial J}
It is important to note that Eqs. \ref{eq.JLWIII} and \ref{eq.JLWII} are valid for a mean, uniform LW background. However, it is possible that a halo would have some stellar sources in neighbouring haloes which would produce levels of $J_{\rm LW}$ higher than the global mean value, which would depend on the clustering scale of haloes \citep[KA09 hereafter]{Ahn:2009p77}.

In order to calculate the effects of a spatially varying Lyman Werner specific intensity from individual Pop III stars, we write 
\begin{eqnarray}
J^{\rm III}_{\rm local} = \frac{f_{\rm esc}}{\pi}\frac{h\nu_{\rm avg}}{\Delta \nu_{\rm LW }}\frac{{Q_{\rm LW}}}{4 \pi d^2} \ ,
\label{eq.JspIII}
\end{eqnarray}
were $h\nu_{\rm avg}$ is the average energy of a photon emitted from a Pop III star in the LW band, $\Delta \nu_{\rm LW}$ is the difference in the maximum and minimum value of the LW frequency range, $d$ is the luminosity distance, and $Q_{\rm LW}$ (expressed as; $Q_{\rm LW} = Q_{\rm H_2} - Q_{\rm H}$) is the number of photons produced per second in the LW energy range. The factor of $\pi$ in Eq. \ref{eq.JspIII} $\left(\frac{f_{\rm esc}}{\pi}\right)$ arises from the conversion of the flux into specific intensity, assuming that each Pop III star is a uniform bright sphere \citep{Rybicki:1986p1295}. The specific values that we use for these parameters in the case of Pop III stars have been computed using the functional fits from \cite{Schaerer:2002p21} where they  track the evolution of stars with different masses and metallicities in their models. They express a given parameter $X$ (see Table \ref{table.func fits}) as a function of the stellar mass $M_*$, as follows: 
\begin{eqnarray}
\log (X) = a_0 + a_1m +a_2m^2 + a_3m^3 \ ,
\label{eq.sch func}
\end{eqnarray}
where $m=\log (\frac {M_*}{\ \rm M_{\sun}})$.

For the contribution of LW photons from Pop II stellar sources, since we only form a total mass in Pop II stars ($M_{\rm *,II}$, see section \ref{sec:popii}), we calculate Pop II properties using the data\footnote{
The data from \textsc{starburst99}, Fig. 7e assumes a Salpeter IMF with a mass cut off at $1,100 \ \rm M_{\sun}$, instantaneous star formation, total stellar mass $=10^6 \ \rm M_{\sun}$, metallicity of $Z=0.001$ and no nebular emission. These parameters are the closest to a Pop II stellar population.} obtained from the \textsc{starburst99} catalogue. 
We integrated the curve(s), in the LW range to obtain a function $\dot{E}$  which is the energy per unit time (in units of erg sec$^{-1}$) emitted by a $10^6 \ \rm M_{\sun}$ Pop II stellar population as a function of the age, as shown in Fig. \ref{fig.ST99}. We then calculate 
\begin{eqnarray}
J^{\rm II}_{\rm local} = \frac{f_{\rm esc}}{\pi}\frac{\dot{E}}{\Delta\nu_{\rm LW}}\frac{M_{6,\rm *,II}}{4\pi d^2} \ ,
\label{eq.JspII}
\end{eqnarray}
where $M_{6,\rm *,II}$ is the mass of the stellar population normalised to $10^6 \ \rm M_{\sun}$. Similar to Eq. \ref{eq.JspIII}, the factor of $\pi$ in Eq. \ref{eq.JspII} $\left(\frac{f_{\rm esc}}{\pi}\right)$ arises from the conversion of the flux into specific intensity, assuming that the Pop II stellar population is a uniform bright sphere \citep{Rybicki:1986p1295}. We form Pop II stars following the prescriptions described in Sec. \ref{sec:popii}. 
\begin{figure}
\includegraphics[width=0.47\textwidth,height=0.34\textwidth,trim=15mm 0 0 0]{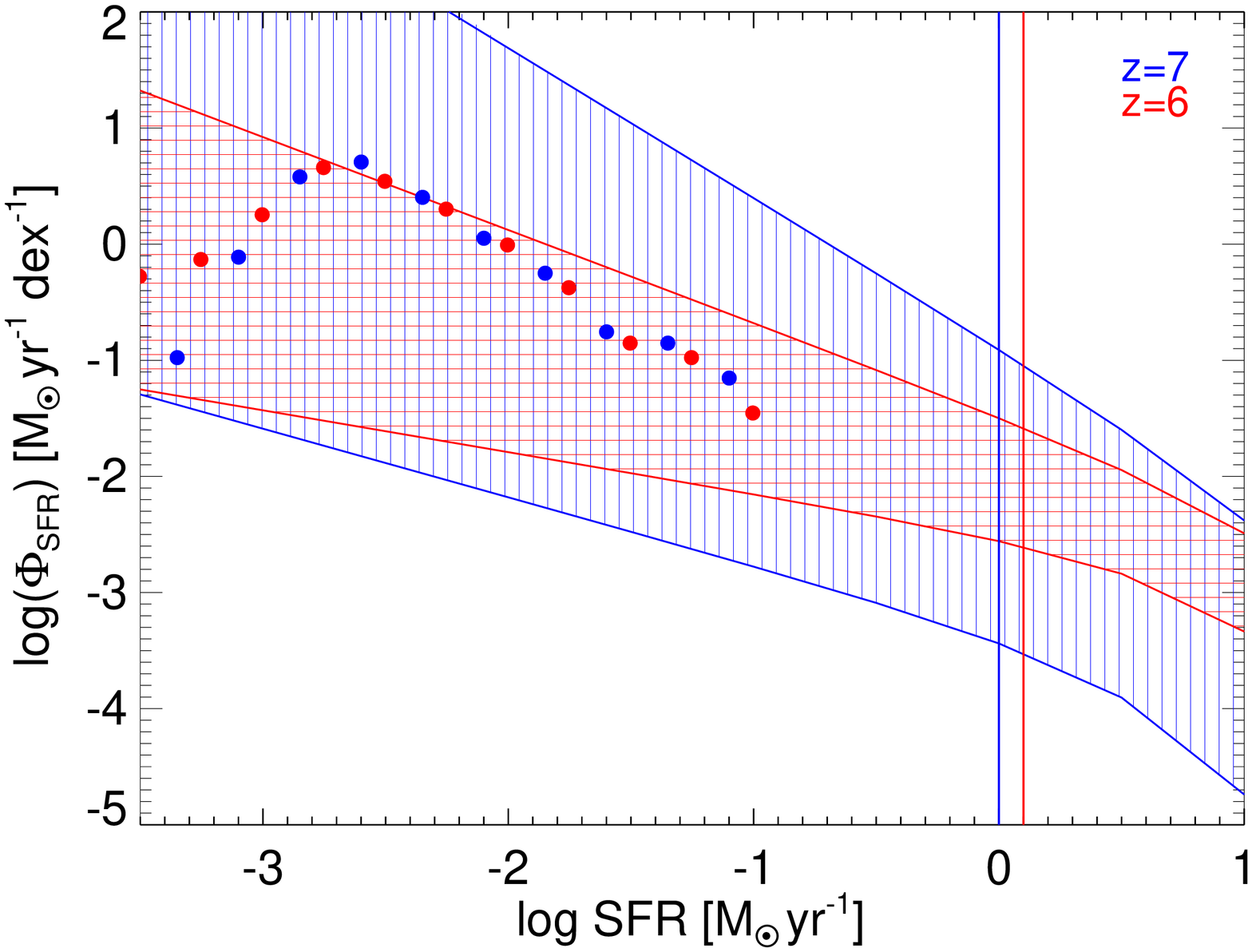}
\caption{Comparison between the SFR-function as computed from S12 and our work. The $\Phi_{\rm SFR}$ bounds at $z=6$ and $7$ as computed from S12 are represented by the red and blue regions whereas the data from our simulation is marked as the red ($z=6$) and blue ($z=7$) filled circles. The current observational surveys are able to probe the SFRs rightwards of the vertical lines denoting $\rm SFR \sim 1 M_{\sun}\ \rm yr^{-1}$ \protect{\citep[e.g.][]{Smit:2012p1007}}.}
\label{fig.smitmf}
\end{figure}
As discussed in Section \ref{sec:MeanJ}, we assume each halo is exposed to a minimum level of $ J = J_{\rm bg}$ calculated in each of our runs at each snapshot. We then add up the LW contribution from each star on top of the background to get the total value of the LW radiation that a halo is exposed to. This slightly overestimates the LW contribution by a factor of less than a few percent. However as we will show later this will not impact our results. To summarise, we have
\begin{eqnarray}
\label{eq.jiii}
J^{\rm III}   & =& J^{\rm III}_{\rm bg} + J^{\rm III}_{\rm local} \ ,\\
\label{eq.jii}
J^{\rm II}    & =& J^{\rm II}_{\rm bg} + J^{\rm II}_{\rm local} \ ,\\
\label{eq.jadd}
J_{\rm total} & =&J^{\rm III} + J^{\rm II} \ .
\end{eqnarray}
The quantity $J_{\rm total}$ is only used for determining if the pristine minihaloes can host Pop III stars (see Fig. \ref{osheanorman}). In their work, \cite{OShea:2008p41} analysed the gas collapse within haloes in the presence of a $J_{\rm LW}$ flux. The photons could be coming from Pop II, Pop III or both as long as the photons are in the correct energy band, hence Eq. \ref{eq.jadd} is valid for analysing pristine minihaloes for Pop III star formation.

On the other hand, due the importance of the spectrum at lower energies for the dissociation of H$^{-}$, the quantities $J^{\rm III}$ and $J^{\rm II}$ are used to determine if the gas in the halo can undergo DC by comparing the values to $J^{\rm III}_{\rm crit}$ and $J^{\rm II}_{\rm crit}$ respectively.
\subsection{Escape fraction of LW radiation and reionization feedback }
\begin{figure}
\centerline{ \includegraphics[width=0.5\textwidth,height=0.34\textwidth]{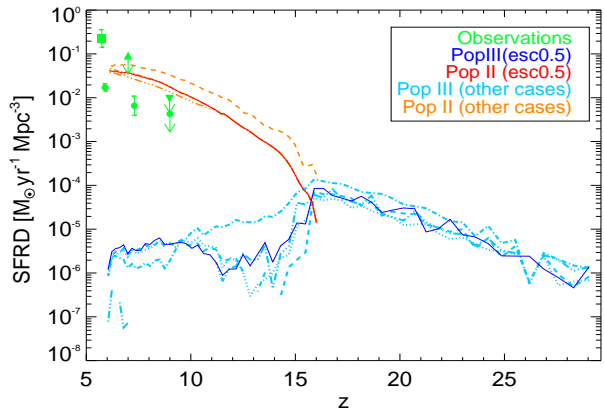} }
\caption{SFRD computed using the methods described in Sec. \ref{sec:Method} for all the cases in our work (see Table. \ref{tablecases}). Solid, dotted, dashed, dash-dot-dash and dash-double-dot-dash represent the cases esc0.5, esc1.0, esc0.5HSFE, esc0.1 and esc0.5reion respectively. The green square, upright triangle, inverted triangle and circles represent observational data from \citet{Hopkins:2004p977}, \citet{Mannucci:2007p978}, \citet{Bouwens:2008p983} and \citet{Laporte:2012p928} respectively.}
\label{fig:sfrI}
\end{figure}
Recent studies \citep{Wise:2009p455,Yajima:2011p1296,Paardekooper:2011p1297} have shown that that the escape fraction for UV photons could vary with the parent halo mass, however, the precise values of LW escape fractions from haloes is still unclear. One might argue that once a pristine halo has hosted a Pop III star (or even a Pop II stellar cluster), most of the H$_{\rm 2}$ is depleted in the halo and the LW photons should, in principle, escape the halo unobstructed, implying $f_{\rm esc, halo} \simeq 1.0$ (KA09). However, the stars (which form in dense environments within the halo) are expected to be surrounded by molecular hydrogen, hence implying a $f_{\rm esc, halo}<1.0$. Previous studies , \citep{Ricotti:2001p1122, Kitayama:2004p669} have found that the minimum escape fraction for LW photons can be $0.1$ but can also reach values of $\ga$ 0.8 in minihaloes.

In addition to the $f_{\rm esc, halo}$, the optical depth, $\tau_{\rm LW}$, of the inter-galactic medium (IGM) would also impact the number of LW photons reaching a neighbouring halo. \cite{Ciardi:2000p82} find that, typically, $\tau_{\rm LW} \ltsim 3$, and including this in our calculations would imply an additional factor of $e^{-\tau_{\rm LW}}$ in Eqs. \ref{eq.JLWIII}, \ref{eq.JLWII}, \ref{eq.JspIII} and \ref{eq.JspII}. Note that effectively, the $f_{\rm esc}$ used in our work can be viewed as a degenerate combination of an escape fraction of LW photons from the halo, $f_{\rm esc,halo}$, and the optical depth of the IGM i.e. $f_{\rm esc} = f_{\rm esc,halo} \times e^{-\tau_{\rm LW}}$. 

Given the uncertainty in $f_{\rm esc, halo}$ and $\tau_{\rm LW}$, we chose three cases to bracket the range of possibilities: $f_{\rm esc}=0.1, 0.5, 1.0$. We also ran a case for our model in  which we implemented an additional reionisation feedback from hydrogen-ionising photons by setting a circular velocity threshold of $20\ \rm km\, s^{-1}$ for all the haloes with $T_{\rm vir}>10^4\ \rm K$ at $6<z<10$. This choice is motivated by the work of \cite{Dijkstra:2004p775} where they study the gas collapse in haloes under a photo-ionising flux and find that a halo must be above a certain mass threshold (characterised by circular velocity in their work) to allow for at least half of the gas to undergo collapse. Other studies \citep[e.g.][]{Okamoto:2008p828,Mesinger:2008p459} have also looked into the feedback process and have found similar mass thresholds.
\subsection{Model normalisation }
For our fiducial case, esc0.5, we set $f_{\rm esc}$$=$$0.5$ for the LW radiation, $\alpha$$=$$0.005$ and implement LW feedback in the model. The choice of $\alpha$ is made in order to match the observations of the cosmic SFRD at $z\gtsim6$. We normalise our free model parameter for the star formation efficiency against recent observations of the SFRD \citep{Hopkins:2004p977, Mannucci:2007p978, Bouwens:2008p983, Laporte:2012p928}. Due to the sensitivity limits of present surveys, the range in star formation rates probed in our simulations is not observed. Thus we chose the fiducial value of $\alpha$ in our model to lie within the error limits of the extrapolated faint-end slope of the SFR-function, $\Phi_{\rm SFR}$, of star forming galaxies at $z=6$ and $7$ as shown in \cite{Smit:2012p1007} (S12 hereafter). In Fig. \ref{fig.smitmf}, the red and blue regions enclose the limits on $\Phi_{\rm SFR}$ at $z=6$ and $7$ respectively, constructed using the fit parameters provided in S12. The blue and red filled circles denote the data from our work which is in fair agreement with the expected values of $\Phi_{\rm SFR}$.

We vary $f_{\rm esc}$ to $0.1,1.0$, keeping $\alpha$ constant at $0.005$ and name the cases esc0.1 and esc0.5 respectively. The model labelled esc0.5Reion is where we also account for reionisation feedback effects on top of the fiducial case. We also implemented a high star formation efficiency for Pop II stars by setting $\alpha$$=$$0.1$ and label the model as esc0.5HSFE. All our cases are summarised in Table \ref{tablecases}.  

\section{Results}
\label{sec:Results}
\begin{figure}
\centerline{\includegraphics[width=0.5\textwidth,height=0.34\textwidth]{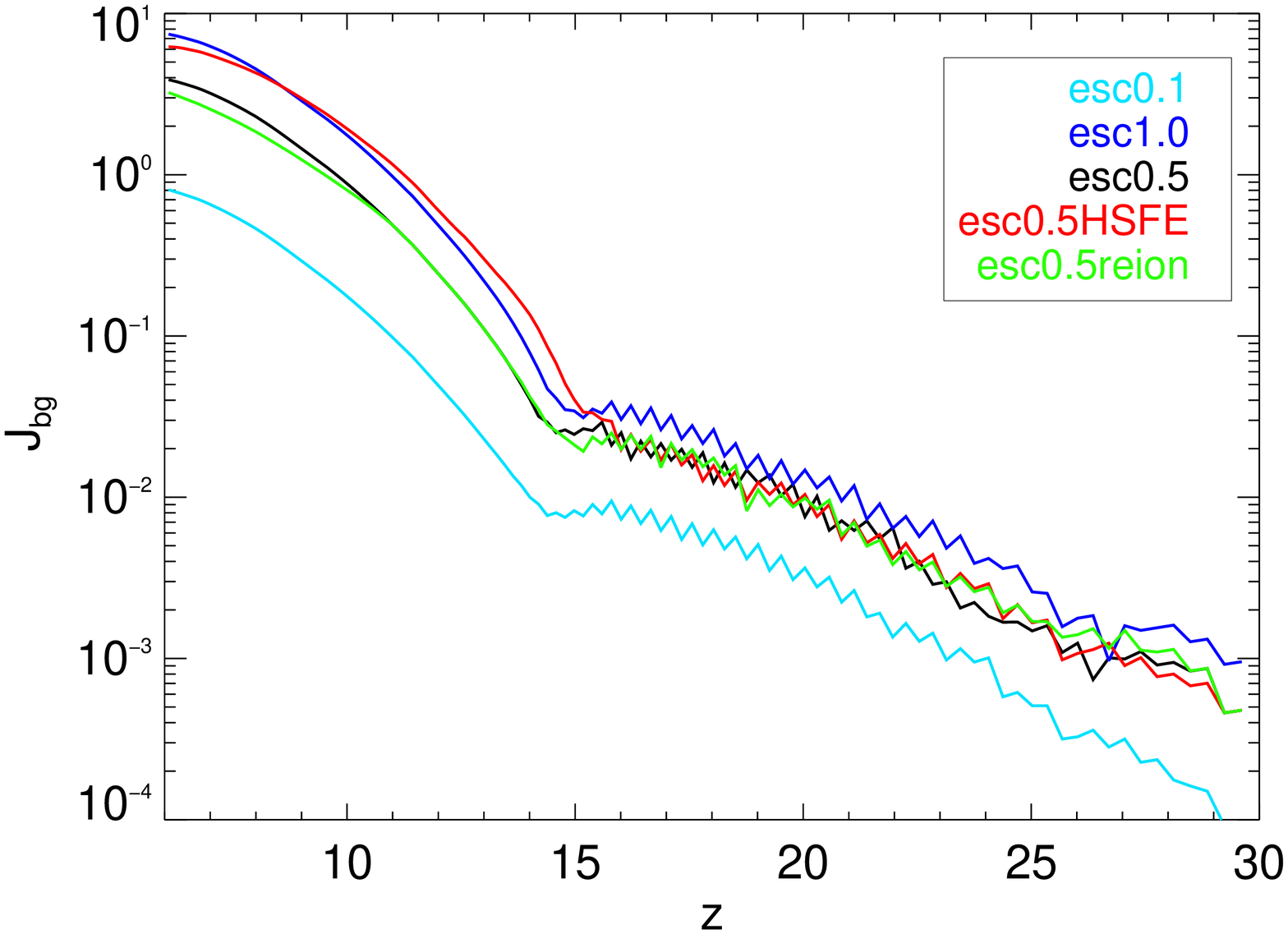}} 
\caption{The self consistent build up of $J_{\rm bg}$, defined as the addition of the LW backgroind radiation from both the stellar populations, plotted against redshift. In all the cases except esc0.5reion, higher $J_{\rm bg}$ leads to a higher number of total DCBHs due to the efficient LW feedback and the resulting higher $J_{\rm total}$. }
\label{fig.jnetmean}
\end{figure}
%
\begin{table}
\caption{Summary of cases considered in our work. The fiducial case is marked in bold.}
\begin{center}
\begin{tabular*}{0.45\textwidth}{@{\extracolsep{\fill}}llll}
\hline
Case & $f_{\rm esc}$ & $ \rm Feedback$ & $\alpha$ \\ 
\\
\hline
esc0.1 & 0.1 & LW & 0.005 \\ 
esc1.0 & 1.0 & LW & 0.005 \\ 
\bf{esc0.5} & \bf{0.5} & \bf{\rm LW} & \bf{0.005}\\ 
esc0.5HSFE & 0.5 & LW & 0.1 \\ 
esc0.5Reion & 0.5 & LW + reionisation & 0.005 \\ 

\hline
\end{tabular*}
\end{center}
\label{tablecases}
\end{table}
The evolution of the SFR density (SFRD) with redshift for all cases is plotted in Fig. \ref{fig:sfrI}. The green symbols represent the cosmic SFRD as inferred from observations. The SFRD we compute is within the observational constraints at $z\sim6$. Pop II stars first appear in our box at $z \sim 16$. The red and blue solid lines in the plot represent our fiducial case of esc0.5 and the light blue and orange lines represent the other cases in our work. The Pop II SFR is roughly the same in all the cases except for esc0.5HSFE and esc0.5Reion. However, due to the different escape fractions assumed for the cases, the Pop III star formation varies over all redshifts. This is due to the fact that in our model, although the Pop III star formation is critically affected by the self consistent build up of the LW radiation, Pop II star formation is not. In general, the Pop III SFRD is inversely proportional to the number of LW photons produced, which is directly proportional to the escape fraction. This is illustrated by the higher level of the Pop III SFRD in the esc0.1 case (light blue, dash-dot-dash line) than all the others.

After $z \sim16$, the increase in the LW radiation due to the Pop II stars (see the following sections) is able to further suppress the Pop III star formation. Also, maximal suppression of Pop III star formation is observed in the esc0.5Reion case where the additional mass constraint of $V_{\rm c}=20 \ \rm km\, s^{-1}$ between $6<z<11$ prohibits the pristine minihaloes and massive haloes from making Pop III stars. The additional circular velocity threshold in the esc0.5Reion case also causes a drop in the Pop II SFRD at $z<11$ which is again due to the fact that the halo mass corresponding to $V_{\rm c} \ge 20 \ \rm km\, s^{-1}$, the assumed mass threshold for structure formation in esc0.5Reion, is slightly higher than $10^8 \ \rm M_{\sun}$ which is the mass threshold for Pop II star formation in all the other cases. 
\subsection{The LW intensity}
We start by expressing the total mean LW intensity at a snapshot as the sum of the contribution from the Pop III and Pop II stellar sources. 
\begin{equation}
J_{\rm bg}=J^{\rm III}_{\rm bg} +J^{\rm II}_{\rm bg} \ .
\end{equation}
The evolution of $J_{\rm bg}$ for all our cases is plotted in Fig. \ref{fig.jnetmean}. Note that in each case, $J_{\rm bg}$ scales as the product of the escape fraction and SFRD at a given redshift. This is illustrated by the fact that although the level of the Pop II SFRD for esc0.5 and esc0.1 is similar, as seen in Fig. \ref{fig:sfrI}, the level of $J_{\rm bg}$ is lower for esc0.1 than esc0.5 in Fig. \ref{fig.jnetmean}. The higher level of $J_{\rm bg}$ for esc0.5HSFE than esc1.0 for $11<z<16$ can be explained by the SFRD in the respective cases. The SFRD is considerably higher between $11<z<16$ for esc0.5HSFE, however once the SFRD approaches that of esc1.0 (not visible in Fig. \ref{fig:sfrI} as it is hidden by the red line), the esc1.0 case produces more LW photons and hence a higher value of $J_{\rm bg}$ is seen for esc1.0 at $z<11$ as compared to esc0.5HSFE. The build up of $J_{\rm bg}$ is in sync with the the SFRD in each of the cases and hence, consistent with our implementation.
\begin{figure*}
\centerline{\includegraphics[width=0.8\textwidth,height=0.5\textwidth]{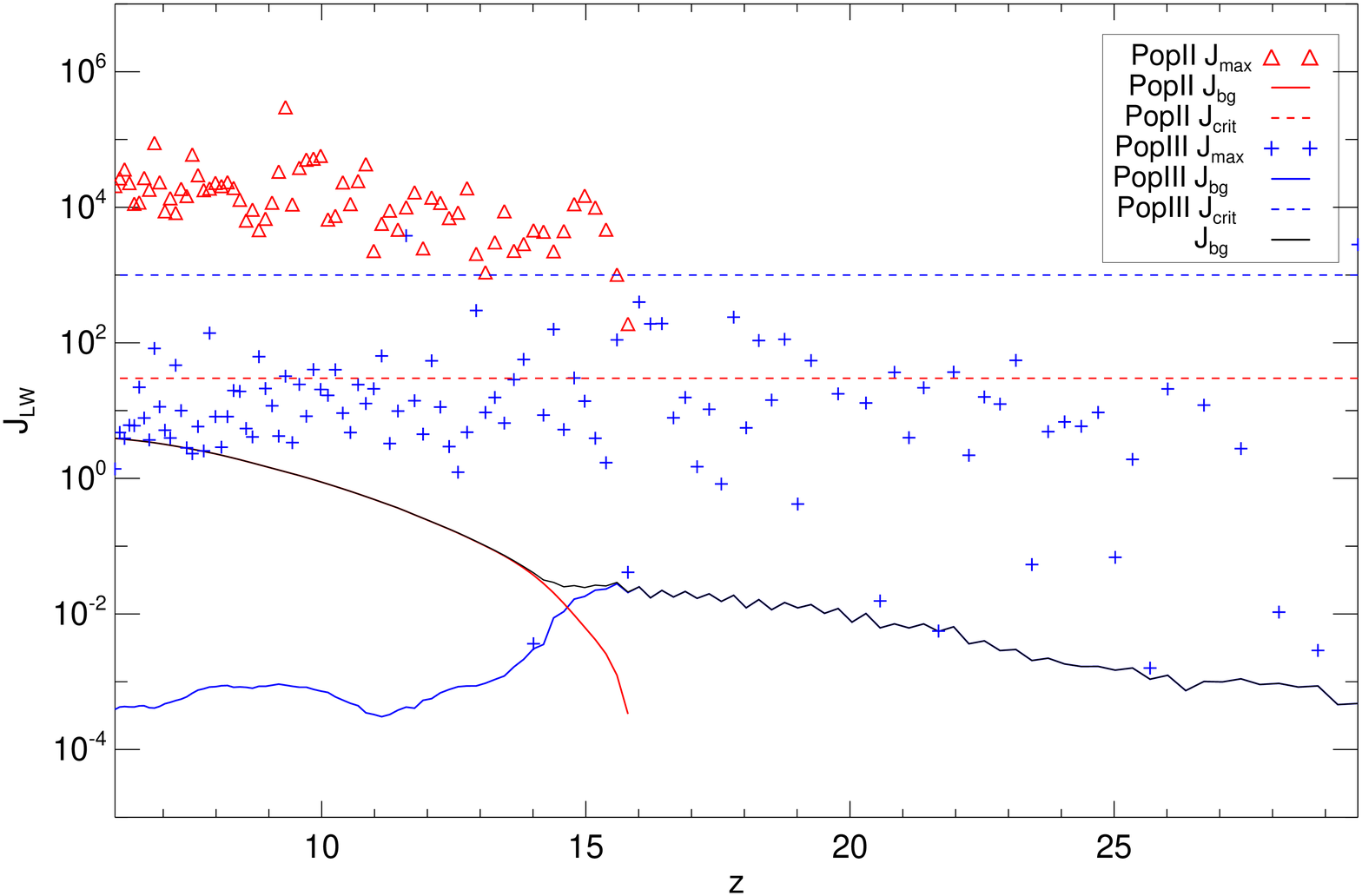}}
\caption{The mean and maximum level of LW radiation plotted for each redshift in our fiducial case. The red triangles ($J^{\rm II}$) and blue crosses ($J^{\rm III}$) indicate the maximum value of LW radiation to which a pristine halo is exposed at each redshift in the simulated volume. The red and blue dashed lines represent $ J^{\rm II}_{\rm crit}$ and $J^{\rm III}_{\rm crit}$ respectively. It is interesting to see that the maximum value of $J^{\rm III}$ (blue crosses) falls short of $J^{\rm III}_{\rm crit}$ (blue dashed line). However, in the case of Pop II sources, the maximum value of $J^{\rm II}$ (red triangles) is several orders of magnitude higher than the $J^{\rm II}_{\rm crit}$ (red dashed line). Hence we deduce that in our simulation, Pop II sources are most likely the ones to produce the $J_{\rm crit}$  required for direct collapse. Note that the spatial LW radiation is computed only in the pristine haloes with $T_{\rm vir}>2000 \ \rm K$.}
\label{fig.jcomp}
\end{figure*}
In Fig. \ref{fig.jcomp}, we plot the individual backgrounds from both stellar populations and the maximum level of the LW flux seen in a pristine halo at each redshift, for our fiducial case. The solid blue and red lines represent $J^{\rm III}_{\rm bg}$ and $J^{\rm II}_{\rm bg}$ respectively which add up to the solid black line, $J_{\rm bg}$. The $J_{\rm bg}$ computed at each timestep is assumed to be the minimum level of LW radiation to which a halo is exposed at that timestep. As expected, the maximum local level of the LW flux for a stellar population is always higher than the background (and in some rare cases equal to the background level).
At all redshifts (except two cases at $z \approx 12,30$ where the $J_{\rm crit}^{\rm III}$ is seen by minihaloes), Pop III stars produce a $J_{\rm III}$$<$$ J^{\rm III}_{\rm crit}$. On the other hand, Pop II stars are able to produce a  $J_{\rm II}$$>$$ J^{\rm II}_{\rm crit}$ in at least one of the pristine haloes at all redshifts, which is shown by the the red triangles being above the red dashed line. The epoch of DCBH formation in each of our cases is only observed after the Pop II star formation kicks in.

\begin{figure}
\leftline{\includegraphics[width=0.5\textwidth,height=0.34\textwidth]{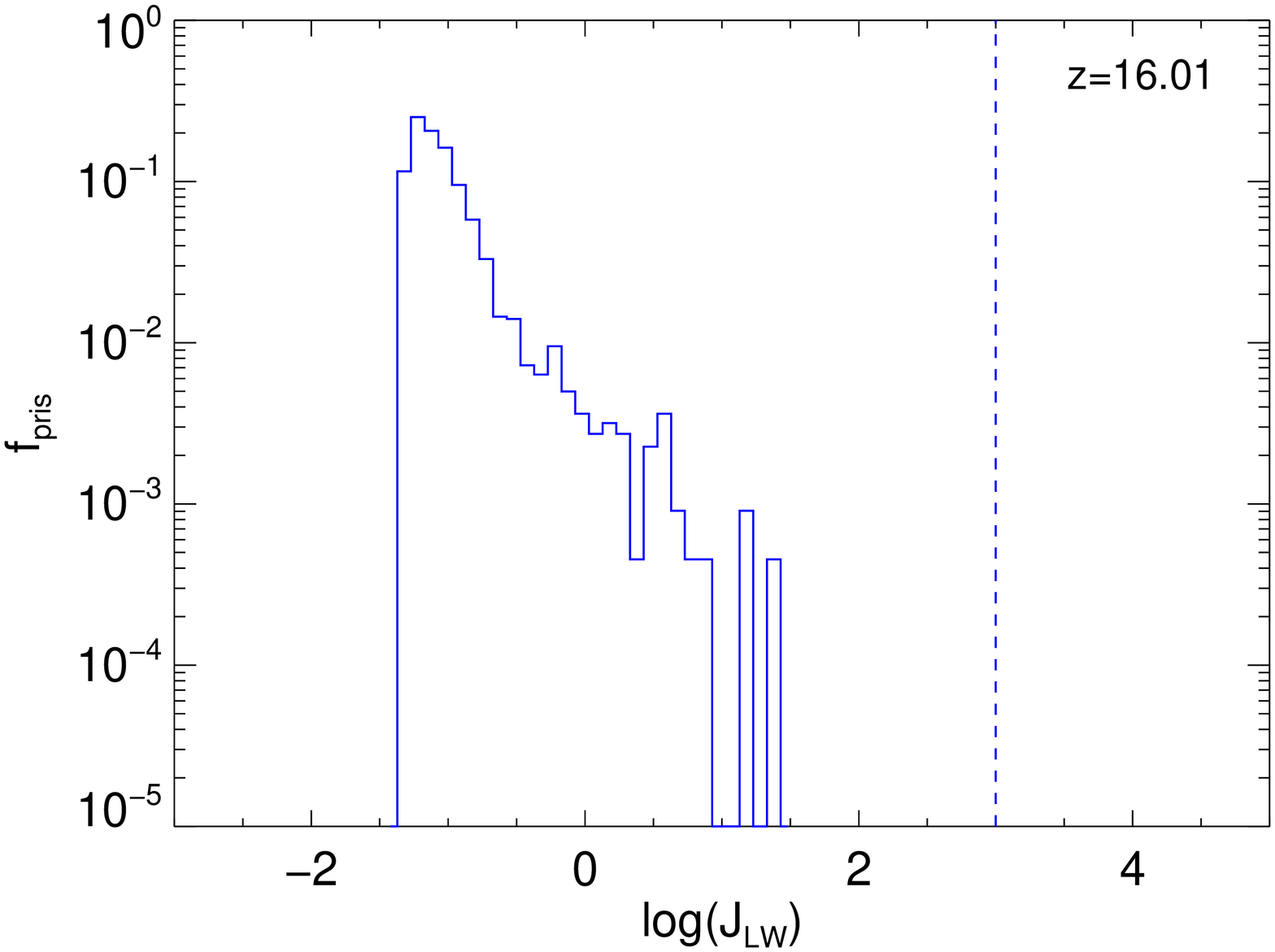}} 
\leftline{\includegraphics[width=0.5\textwidth,height=0.34\textwidth]{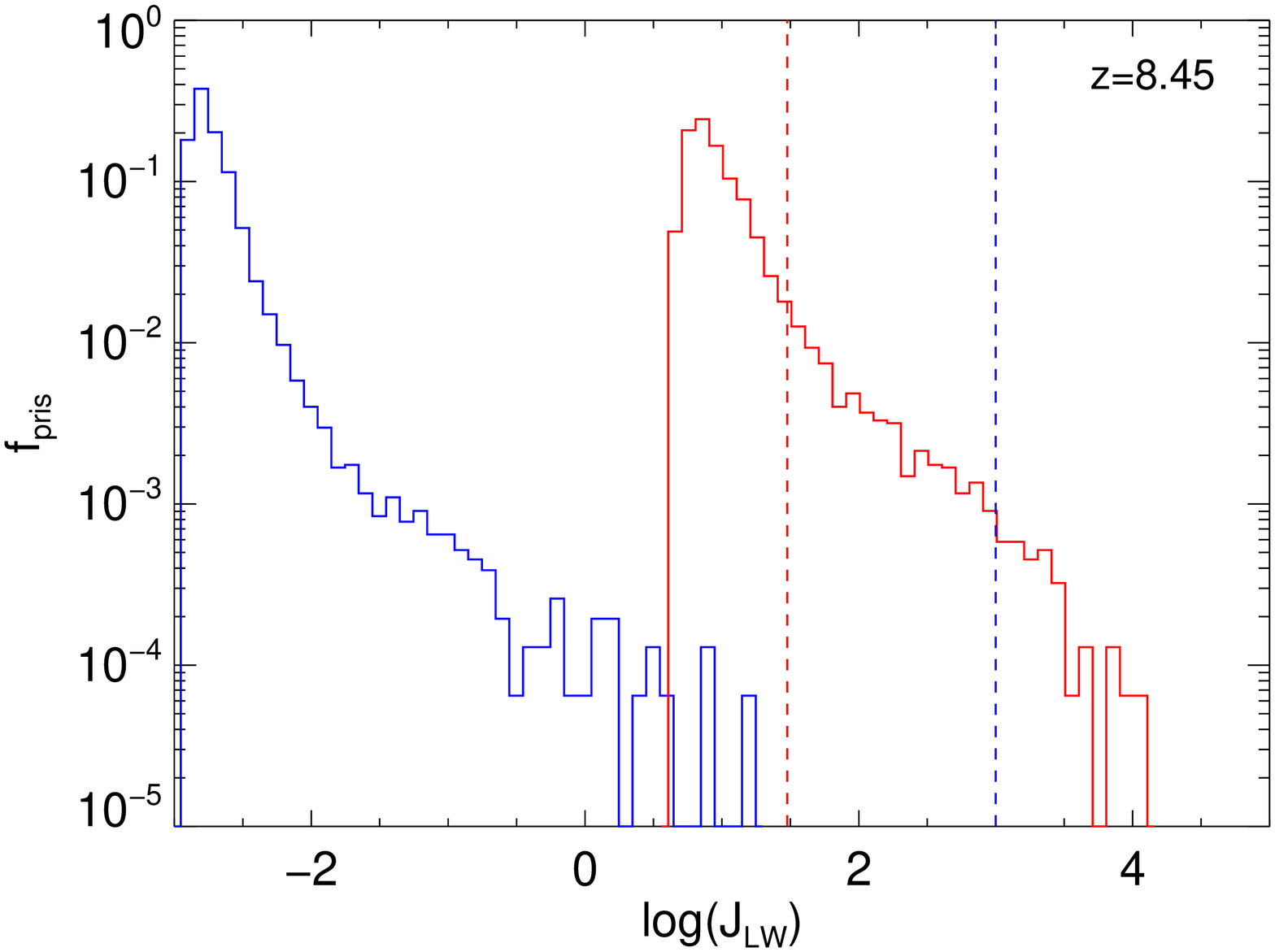}}
\caption{The distribution of $J_{\rm local}$ for all the pristine haloes at a given redshift for our fiducial case. The red and blue solid histograms (dashed lines) indicate $J_{\rm II}$ ($J^{\rm crit}_{\rm II}$) and $J_{\rm III}$ ($J^{\rm crit}_{\rm III}$) respectively. The absence of the red histogram in the top panel is because Pop II star formation begins at $z=16$ in our box. }
\label{fig.jdist}
\end{figure}

We plot the distribution of the local $J_{\rm LW}$ as seen by pristine haloes in Fig. \ref{fig.jdist} before and after the Pop II star formation begins at $z\sim16$. We define $f_{\rm pris}$ as the fraction of pristine haloes with $T_{\rm vir}>2000 \ \rm K$ exposed to a given $J_{\rm LW}$. The red and blue solid histograms represent $J^{\rm II}$ (Eq.\ref{eq.jii}) and $J^{\rm III} $ (Eq.\ref{eq.jiii}) respectively. In Fig. \ref{fig.jdist}, we see that less than one percent of pristine haloes (which roughly translates into a fraction of $5\times 10^{-4}$ of the total number of haloes) see a $J^{\rm II} > J^{\rm II}_{\rm crit}$ whereas the Pop III LW flux is always subcritical even before the Pop II star formation begins. The low fraction of pristine haloes that are exposed to $J^{\rm II}_{\rm crit}$ can be attributed to the rarity of the event where a pristine halo is clustered (hence close enough) to the neighbouring haloes hosting Pop II stars. The trend of the distribution function is similar to D08, where they plot the PDF of all the haloes exposed to varying levels of $J_{\rm LW}$. The Fig. \ref{fig.jdist} further supports our result from Fig. \ref{fig.jcomp} i.e. Pop III stars are always subcritical to DCBH formation and that the LW radiation required for DCBH formation is always produced by Pop II stars.

The value of $J_{\rm bg}$ that we compute in our work for esc0.5 and esc0.5Reion is within $5\%$ of the previous estimates of \cite{Greif:2006p99} at $z = 10$, where they self-consistently study the impact of two types of stellar populations and their feedback on star formation at $z \gtsim 5$. The value of $J_{\rm bg}^{\rm III}$ that we find for all our cases, denoted by the solid lines at $z \gtsim 16$ in Fig. \ref{fig.jnetmean}, is also consistent with \cite{Johnson:2008p29}, i.e. it does not exceed their value of the maximum level of the LW background expected from Pop III stars ($\sim 0.13$ at $z \sim 16$). We also find a good agreement with \cite{Trenti:2009p92} as our SFRD and $J$ outputs resemble their estimates, but it is difficult to draw exact comparisons as they used an analytical Press-Schetcher modelling and the parameter choices of the studies differ considerably. In cases esc0.5 and esc0.5Reion, we find $J_{\rm bg} \approx 1$ at $z\approx 10$ (also see Fig. \ref{fig.jnetmean}) which is very close to the expected value during the reionization era\footnote{
In a recent attempt to model the DCBH formation, \cite{Petri:2012p1335} find a very high global LW flux of $J_{\rm LW}$ $\simeq$ 1000 at the epoch of reionisation and they argue for even higher levels in spatially clustered regions. Our self-consistent methods to calculate the background and spatial variation of the LW flux show that such a high background is difficult to achieve.}
 \citep{Omukai:2001p128,Bromm:2003p22}. 

In accordance with KA09, we find that the local LW intensity, which can be orders of magnitude higher than the mean LW intensity, is observed in highly clustered regions. This becomes even more evident in Sec. \ref{sec:DCBHenv} where we present the cross-correlation functions of DCBH haloes. Note that although KA09 carried out full radiative transfer cosmological simulations to model the spatial variation of the LW intensity, they lacked the resolution required to study the impact of LW radiation feedback on structure formation (namely star formation in minihaloes) at such high redshifts. 
\subsection{Sources responsible for $J>J_{\rm crit}$}
With the change in the relative fraction of Pop III to Pop II star formation, the relative contribution to the LW background undergoes a change as well.  As seen in Fig. \ref{fig.jnetmean}, \ref{fig.jcomp} and \ref{fig.jdist}, the main contribution within our model comes from Pop II stars. At almost all times the contribution from Pop III stars is subcrititcal for DCBH formation. The critical level of radiation required for direct collapse is always produced by Pop II stars. This can be attributed to the way in which Pop III and Pop II stars form.  We checked the effect of allowing Pop III stars to form as single stars, in binary systems or in a group of 10 stars (which maximises the LW flux from a halo) with varying IMFs, and in all the cases the total number of stars was insufficient in producing the critical LW flux in a neighbouring halo. 

Also, the short lifetime of a Pop III star poses a problem as they reach the end of their lifetime in a few $10^6$~yr as compared to a Pop II stellar population which can actively contribute towards critical levels of LW radiation for up to a few $10^{7}$~yr (see Fig. \ref{fig.ST99}). Hence even if it could produce $J_{\rm crit}$, a Pop III star is less likely to be near a massive pristine halo (and hence contribute towards $J_{\rm crit}$) than a Pop II stellar population. The result is in good agreement with \cite{Inayoshi:2011p181}, where they also argue that for a stellar source to produce the $J_{\rm crit}$, it must be a Pop II/I star cluster or very top heavy Pop III galaxies. As per our current understanding,  Pop III stars form in (at most) groups of a few with masses $\sim$ few tens of solar masses \citep[e.g.][]{Greif:2011p93}. The occurrence of Pop III galaxies at such high redshift is expected to be extremely rare as metal pollution in Pop III hosting haloes is quite fast \citep{Maio:2011p104} and it is highly unlikely that a cluster of these short lived Pop III stars could end up in a galaxy \citep{Johnson:2008p29,Johnson:2010p18}. If a massive, pristine halo is to undergo direct collapse, we would expect it to have an external close by neighbour hosting a Pop II stellar population giving rise to $J_{\rm crit}$.

The variation in the spatial LW intensity we find is also consistent with KA09. Using a full radiative transfer prescription in a cosmological box, they found that LW radiation varies on the clustering scale of sources at high redshifts, but they did not resolve the Pop III star forming minihaloes important for such studies.   

\subsection{Abundance and growth of DCBHs}
\label{sec:bh}
\begin{figure}
\centerline{\includegraphics[width=0.5\textwidth,height=0.735\textwidth]{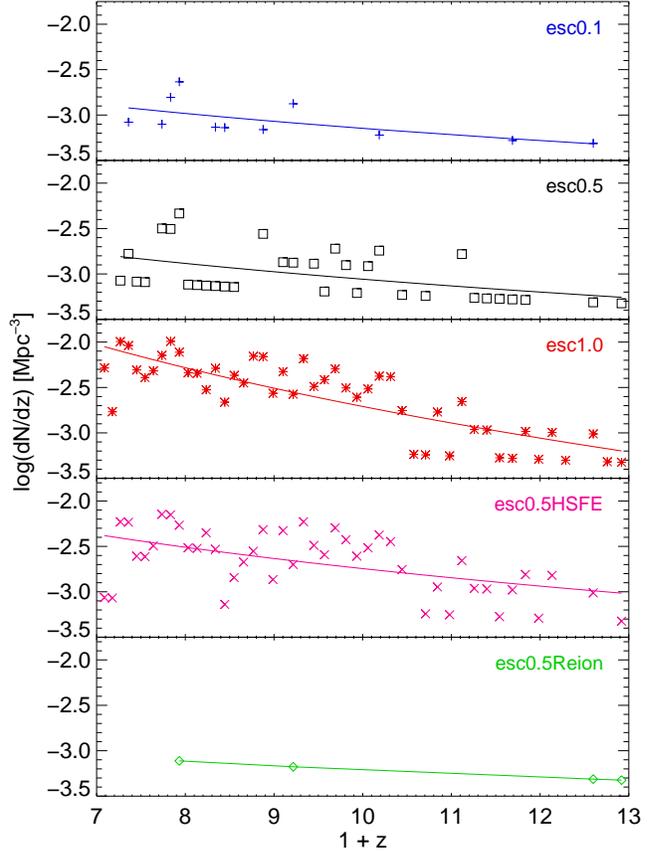}}
\caption{Formation rate density of black holes in all the models plotted against redshift. The line represents a fit (Eq.\ref{dcbhfit}) to the formation rate density in each case.} 
\label{fig.ndens}
\end{figure}

The rate at which DCBHs are found to be forming in our simulation volume is shown in Fig. \ref{fig.ndens}. We find a steady rise in the DCBH formation rate density with decreasing redshift (in units of $\rm Mpc^{-3}$) which can be expressed as
\begin{equation}
\frac{dN}{dz} = b_1\ (1+z)^{b_2} \ ,
\label{dcbhfit}
\end{equation}
where $b_1$, $b_2$ are the fit parameters for the DCBH formation rate in each of our cases as shown in Table \ref{tabledcbhfit}. As we are neglecting possible metal pollution from neighbouring halos \citep{Maio:2011p104} the formation rates are strict upper limits in each of the cases. The fact that we find a few DCBH candidates in our 3.4 Mpc$h^{-1}$ simulation volume implies that the conditions for a DCBH are achievable in the early Universe and many such intermediate mass BH (if not SMBH) should exist at high redshifts. 

The DCBH formation rate increases as a function of the number of LW photons that are emitted from a star forming halo. In esc0.1 we find 13 times fewer DCBHs than in esc0.5, which is due to the lower escape fraction assumed for LW photons in the former case. As the escape fraction increases from $0.5$ to $1.0$, the DCBH formation rate increases considerably. In all the above cases, this can be explained by the effect arising from the change in $J_{\rm total}$ (Eq. \ref{eq.jadd}) which is two fold
\begin{itemize}
\item a higher $J_{\rm total}$ implies that more minihaloes are prevented from Pop III SF due to efficient LW feedback which makes them available for DCBH formation at later times since they are not metal-enriched.
\item a higher $J_{\rm total}$ directly affects the efficiency of DCBH formation as it easier to exceed $J_{\rm crit}$.
\end{itemize}
The lower formation rate of DCBHs in esc0.5HSFE than esc1.0 can be attributed to the lower level of $J_{\rm bg}$ in the former case at $z<11$. Also, since the majority of DCBHs form at $z<11$, the fits are dominated by the DCBH formation rate at later times and hence a lower slope for esc0.5HSFE than esc1.0 is seen.

The esc0.5reion case produces an interesting outcome where we find only 4 DCBHs which is roughly 10 percent of the DCBHs produced in our fiducial case. Note that before the reionisation feedback kicks in at $z=11$, both esc0.5 and esc0.5Reion have the same DCBH formation rate. One would expect photo-ionisation effects and the photo-evaporation of pristine minihaloes in the early Universe, which is accounted for by our reionisation feedback model, to greatly reduce the number of haloes into which primordial gas can collapse at later times. We discuss this case in more detail in Sec. \ref{sec:reion}.

To explore the impact of BH growth via accretion after their formation, we allow the BHs to grow via Eddington accretion using the relation
\begin{equation}
M_{\bullet}(t)=M_{\bullet,0} \ {\rm exp}\left(f_{\rm edd}\frac{1-\epsilon}{\epsilon}\frac{t}{450 \mbox{ Myr}}\right) \ ,
\label{eddacc}
\end{equation}
where $M_{\bullet}$ is the final mass of the black hole, $M_{\bullet,0}$ is the initial mass of the black hole set to $10^4\ \rm M_{\sun}$ for a DCBH \citep[e.g.][CS10]{Johnson:2012p874}, $t$ is is the accretion time, $\epsilon$ is the radiative efficiency and $f_{\rm edd}$ is the Eddington fraction. We explore the $(f_{\rm edd}, \epsilon)$ parameter space by choosing $\epsilon=0.07,0.1,0.2$ to account for a range in radiative efficiencies for Eddington ($f_{\rm edd}=1$), sub-Eddingtion ($f_{\rm edd}<1$) and super-Eddington ($f_{\rm edd}>1$) accretion \citep {Johnson:2011p704,Shapiro:2005p765}. Since there is a lot of ambiguity regarding the early regimes of BH accretion, we varied our Eddington accretion parameters ($f_{\rm edd} = 0.4,1.0,1.5$) to account for a range of possibilities in the overall accretion mode of the BH.

\begin{figure}
\includegraphics[width=0.5\textwidth,height=0.34\textwidth]{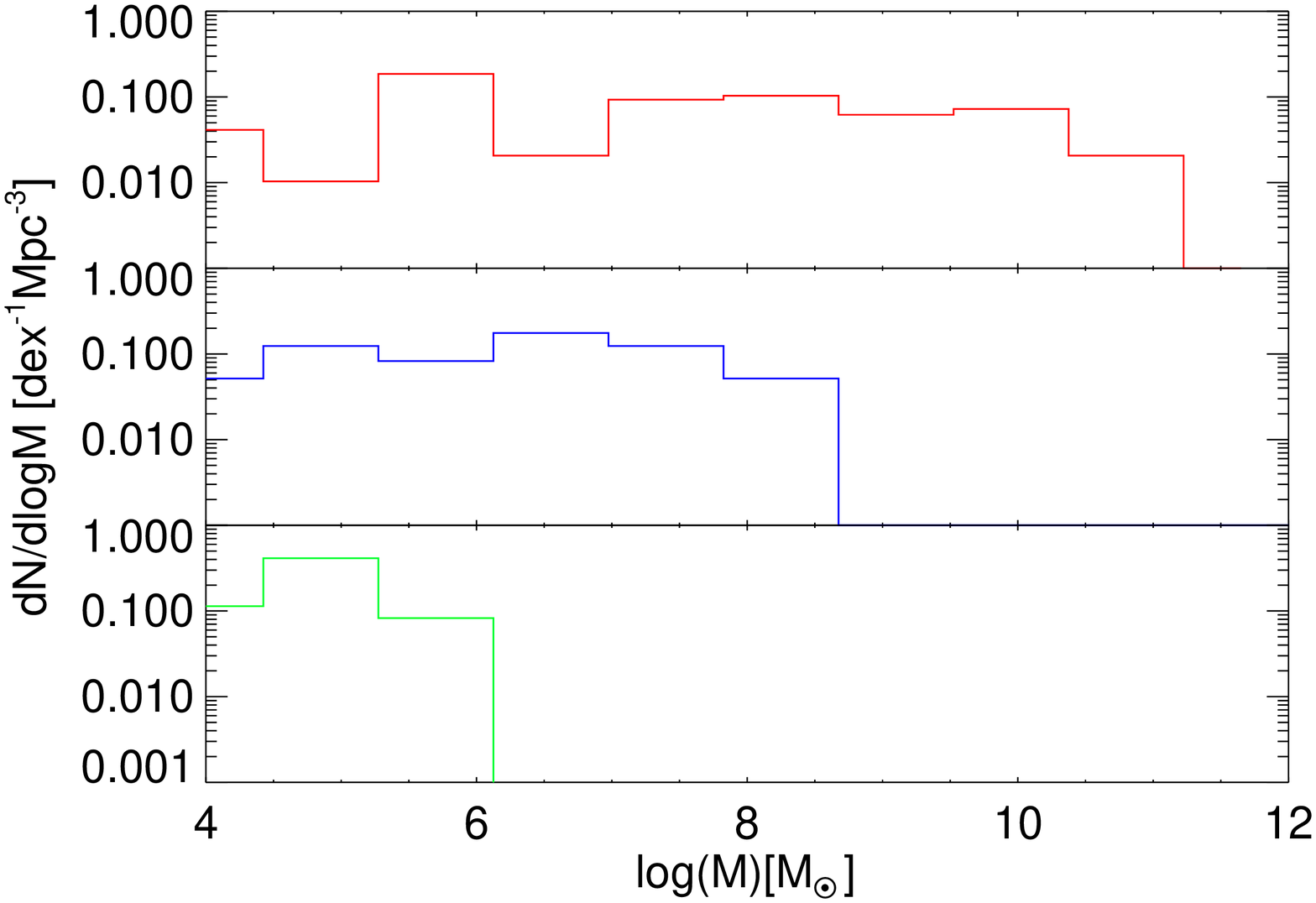}\\
\includegraphics[width=0.5\textwidth,height=0.34\textwidth]{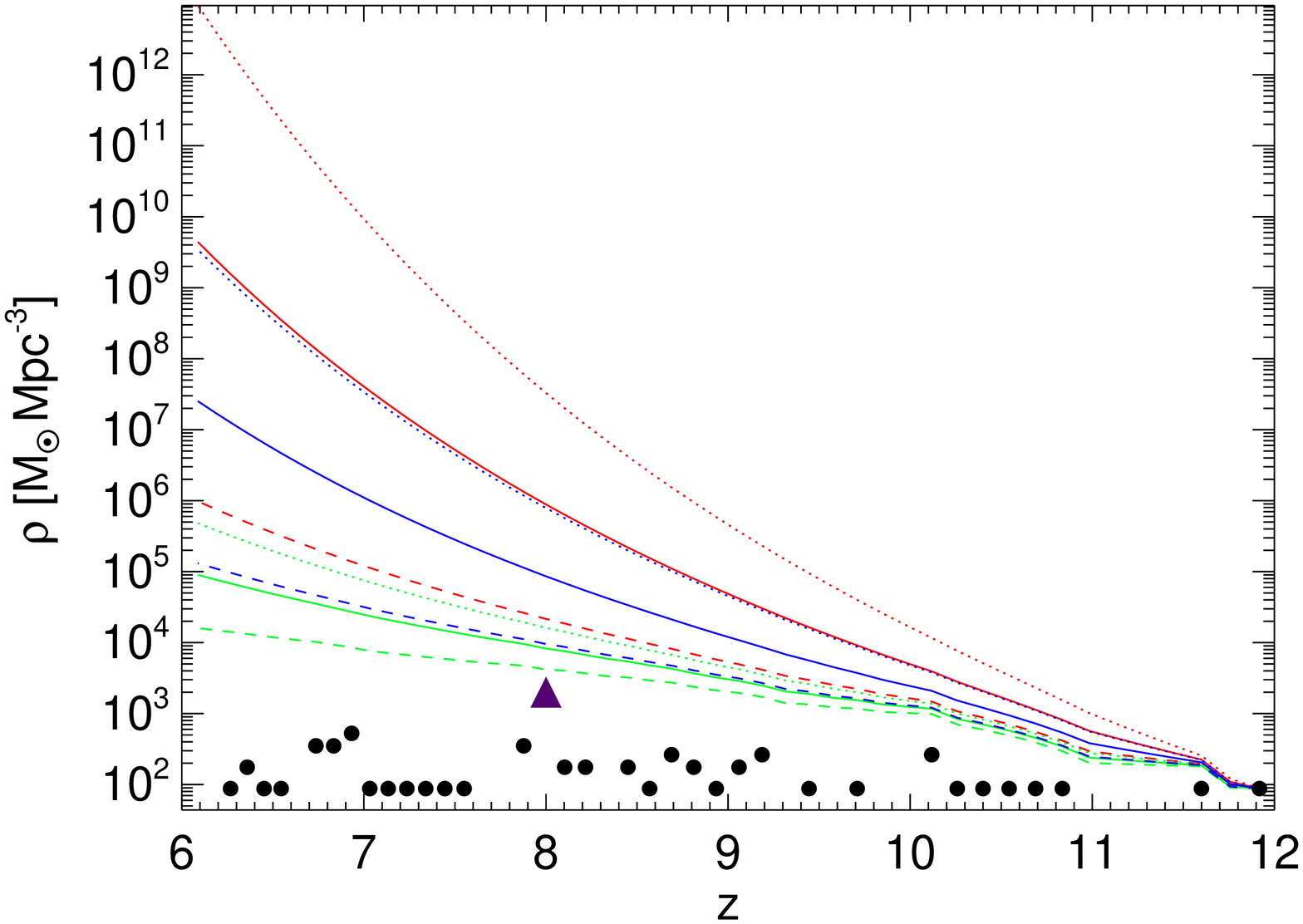}
\caption{DCBH mass function and cumulative mass density for the fiducial case (esc0.5). Red ($f_{\rm edd}=1.5$), blue ($f_{\rm edd}=1$) and green ($f_{\rm edd}=0.4$) represent the super-Eddington, Eddington and sub-Eddington scenarios respectively. \textit{Top Panel:} mass function of accreting DCBHs at $z=6$ for radiative efficiency $\epsilon=0.1$. \textit{Bottom Panel:} cumulative mass density of the accreting DCBHs vs. redshift for $\epsilon=0.07, 0.1, 0.2$ represented by dotted, solid and dashed lines respectively and following the same colour coding as the top plot for $f_{\rm edd}$. The black circless represent the total mass density of the newly formed DCBHs at each redshift. The solid purple triangle marks the claim of T11 at $z\approx8$.}
\label{fig.bhtwin}
\end{figure}

For our fiducial case (esc0.5), Fig. \ref{fig.bhtwin} shows the DCBH mass function constructed for $f_{\rm edd}=0.4,1,1.5$ and $\epsilon=0.1$ (top panel) and the cumulative mass density of these DCBHs at $z=6$ (bottom panel) plotted for different choices of $f_{\rm edd}$ and $\epsilon$. It is clear from the top panel that the DCBHs can almost reach SMBH scales with Eddington accretion and quite easily attain a mass larger than $10^9 \ \rm M_{\rm sun}$ if we assume super Eddington accretion. A wide range of evolved BH mass densities is seen in the bottom panel. Each filled black circle in the bottom panel represents the mass density of newly formed DCBH at that redshift, assuming an initial DCBH mass of $10^4 \ \rm M_{\sun}$. The solid purple triangle in the bottom panel is the observational claim made by T11 for the mass density of the IMBH at $z\sim8$. Although we do not match T11's claim for esc0.5, we do so for esc0.5Reion as explained in section \ref{sec:reion}.

\subsection{DCBH Host Haloes}

In the following sections, we explore the regions where we find the conditions for direct collapse and the histories of the DCBH host haloes.
\begin{table}
\caption{Fit parameters lines in Fig. \ref{fig.ndens} which follow Eq.\ref{dcbhfit} and total DCBH number density at $z=6$ in each of our cases}
\begin{center}
\begin{tabular*}{0.45\textwidth}{@{\extracolsep{\fill}}lccc}
\hline
Case & $b_1$ & $b_2$ & Total DCBH \\
         &             &           &  ($\rm Mpc^{-3}$) \\
\\
\hline
esc0.1 & 0.035 & -1.69&0.13\\
esc0.5 & 0.054 & -1.79 & 0.518\\
esc1.0 & 49 & -4.4 & 2.09\\
esc0.5HSFE & 0.48 & -2.43 & 1.58\\
esc0.5Reion & 0.0061& -1.0 & 0.035\\
\hline
\end{tabular*}
\end{center}
\label{tabledcbhfit}
\end{table}
\subsubsection{Environment}
\label{sec:DCBHenv}

In order to understand the environmental differences between the haloes that host DCBHs and the ones that do not, we construct cross-correlation functions for the distribution of sources around them. In our case, the cross-correlation function is the excess probability of encountering a source in a given distribution of sources around a halo as compared to a uniform distribution.
Consequently, we define the cross correlation function as
\begin{equation}
\xi(d) = \frac{DD(d)}{RR(d)} - 1 \, , 
\label{eq.corr1}
\end{equation}
where $DD(d)$ represents the \textit{data-data} pair counts at a given distance $d$, constructed from our model. The data-data pairs in our work refer to the halo-source pairs. We fix the halo and loop over all the qualifying\footnote{The qualifying sources for a given halo refer to the ones that satisfy the conditions described in section \ref{sec:selection} of the Appendix.} sources thereby computing the physical distances. This is done for all the halo-sources pairs in a given redshift range.

$RR(d)$ represents the \textit{random-random} pair counts constructed from a uniform distribution of sources around a random halo.
Similar to \cite{Li:2012p1265}, we use the formulation of $\xi(d)$ to qualitatively compare the small scale clustering properties of DCBH-hosting and non-DCBH hosting haloes with their respective sources. Note that in our case, $RR(d)$ is the same as $DR(d)$, as used by \cite{Li:2012p1265}, since the position of the halo is arbitrary.

Following the prescription described above, we first construct a cross-correlation function $\xi_{\rm Halo(DC)}$ which is computed over all the newly formed DCBHs, in a given redshift range, and their respective sources. We then define a similar cross-correlation function $\xi_{\rm Halo(NoDC)}$ for haloes that do not host a DCBH with all their respective sources.\footnote{ The non DCBH hosting haloes are chosen in the same mass range as the haloes hosting a DCBH. This is done for consistency in the construction of the cross-correlation function in the two cases.} Both $\xi_{\rm Halo(DC)}$ and $\xi_{\rm Halo(NoDC)}$, are constructed using the same bins at a given snapshot with the first bin placed at a distance larger than the typical virial radius ($\sim 1\ \rm kpc$) of a massive halo. This is done in order to exclude sources that formed within a non DCBH hosting halo at earlier times. This choice does not affect the nature or trend of $\xi_{\rm Halo(NoDC)}$.

\begin{figure}
\centerline{\includegraphics[width=0.5\textwidth,height=0.734\textwidth]{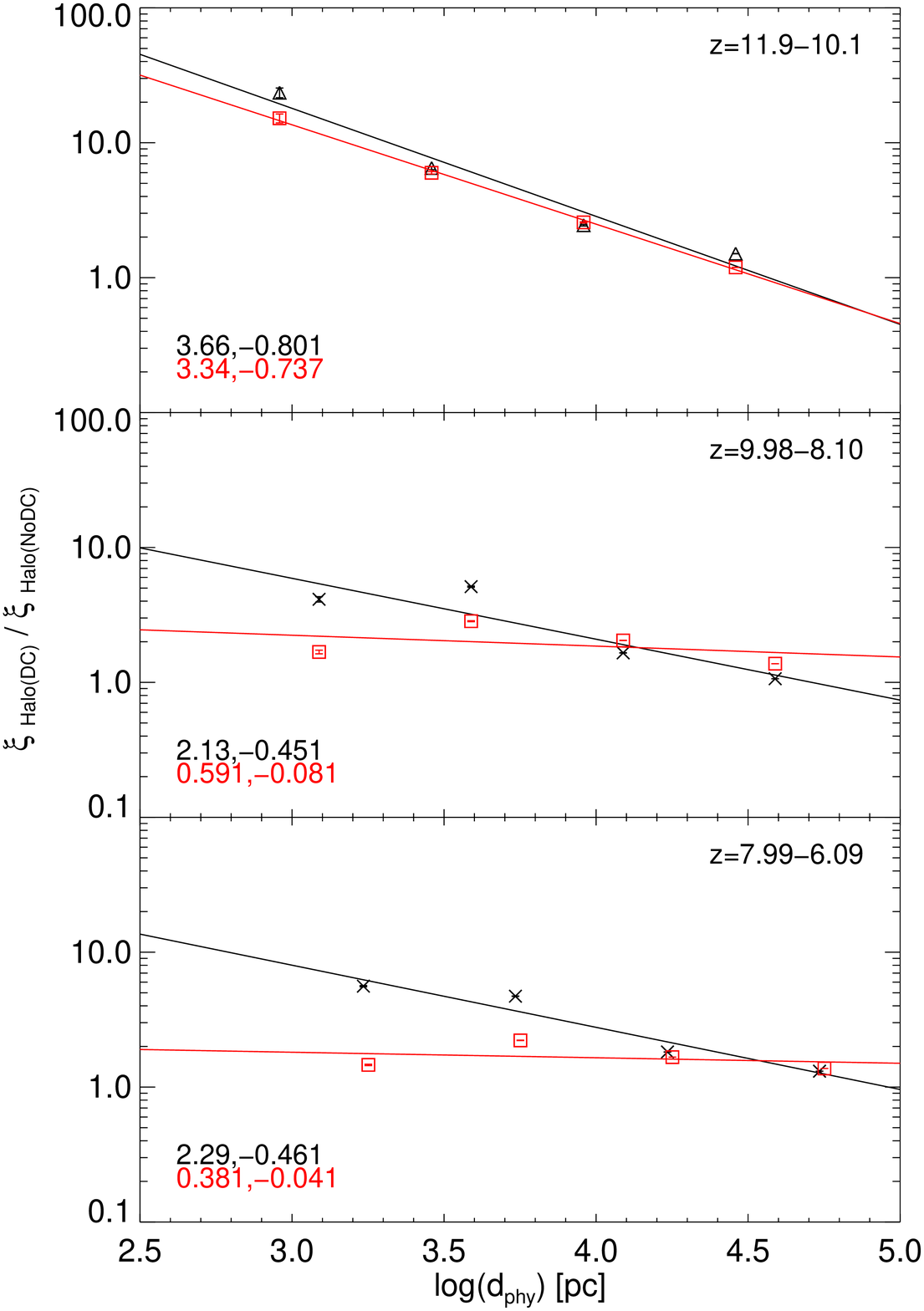}}
\caption{The correlation function $\xi_{\rm total}$ plotted for a range of redshifts (top right of each plot). Black crosses (line) and red squares (line) indicate the correlation data points (fit to the points)  computed from the function $\xi_{\rm total}$ for the cases esc0.5 and esc1.0 respectively. The numbers on the bottom left indicate the fit parameters $c_1$ and $c_2$ in Eq.\ref{eq.fitmain}}
\label{fig.corrdcbh}
\end{figure}

Finally, to check for a variation in the clustering of sources around DCBH hosting haloes versus non-DCBH hosting haloes, we define
\begin{equation}
\xi_{\rm total}=\frac{\xi_{\rm Halo(DC)}}{\xi_{\rm Halo(NoDC)}} \ .
\end{equation}
If $\xi_{\rm Halo(DC)}=\xi_{\rm Halo(NoDC)}$ in each distance bin, it implies that the DCBH host haloes are as clustered as the non-DCBH host haloes at all scales. Therefore, a value of unity for $\xi_{\rm total}$ at a given distance scale would imply the lack of clustering for the DCBH host haloes. Additionally, a variation in $\xi_{\rm total}$ with distance would imply the difference in clustering of the DCBH host haloes vs. non-DCBH host haloes with the neighbouring sources. A negative slope would imply over-clustering for DCBH host haloes at smaller distances whereas a positive slope would imply an over-dense environment at smaller distances for the non-DCBH hosting haloes. The results of our calculations for the cases esc0.5 and esc1.0 are plotted in Fig. \ref{fig.corrdcbh}.\footnote{
The chosen cases have at least a few DCBHs in the specified redshift bins. The case esc0.5Reion and esc0.1 have very few DCBHs and the case 0.5HSFE has not been plotted just to avoid repetition as it lies between the two plotted cases.}

It can be inferred from the value of $\xi_{\rm total}$ and the slope of the fits that the haloes that host a DCBH are more clustered than the non-DCBH hosting haloes especially at a scale of few tens of kpc. This trend can be attributed to the fact that in order to reach $J_{\rm crit}$ near a pristine halo, a source (or a population of sources) must exist very close by.

The different escape fractions make this trend even more prominent since in both the cases, the line gets flatter as we move to lower redshifts but the relative change in the slope of the lines is inversely dependent on the escape fraction. This can be understood by noting that to produce the same level of LW radiation in a neighbouring halo, the sources would need to be closer to the halo (hence more clustered) if the escape fraction was set to 0.5 instead of 1.0. This also implies that the haloes that host DCBH in esc0.5 form from regions of higher over-densities than in esc1.0. Hence, the flattening trend is most pronounced in the esc1.0 (red line) as compared to the esc0.5 (black line).

By using our detailed prescription for the spatial variation of the LW flux, we find that the haloes which are exposed to $J_{\rm crit}$ always have a source within a few kpc. We note that our results are in accordance with previous work done by D08 which shows that the distance scale within which a halo should have a close by LW source in order to undergo direct collapse is $\sim 10 \ \rm kpc$. 

The function $\xi_{\rm total}$ follows a linear fit in log space which can be parameterised as  
\begin{equation}
{\rm log}(\xi_{\rm total}) = c_1 + c_2 \log(d_{\rm phy}) \ ,
\label{eq.fitmain}
\end{equation}
where $d_{\rm phy}$ is the physical distance between a halo and LW source in parsec and the parameters $c_1$ and $c_2$ are indicated on the bottom left of each plot in Fig. \ref{fig.corrdcbh}.

The flattening of the slope can be attributed to the fact that it is more common for a halo to be in an environment with close by sources at later redshifts due to the higher overall SFRD in the box. Also, the haloes that host DCBHs at $z>10$ originate from regions of larger overdensities than the ones at lower redshifts. This is evident from the larger negative slope of $\xi_{\rm total}$ at $z>10$. At lower redshifts, a lower-$\sigma$ fluctuation is required to produce a halo of $\sim 10^7 \ \rm M_{\sun}$, which roughly corresponds to a $T_{\rm vir}=10^4$~K halo, supporting that DCBHs must arise from regions of high over densities in the early Universe.

An important inference can be drawn from the points above. If the DCBHs which form early on (at $z>10$) in the Universe arise from highly clustered regions, their environment can only get more clustered as the halo progresses to later times. This is an important result as it supports the idea that the most massive SMBH that we observe at the centres of ellipticals or as highly clustered AGNs, might have originated from regions of high over-densities quite early on in the Universe, possibly as DCBHs.

\subsubsection{History}
\label{sec.history}

It is interesting to check when the haloes that host DCBHs first originated during their cosmic evolution. We checked the most massive progenitor of each of the DCBHs and tracked it back in time until the halo was found to have a mass equal to the mass resolution in our work (20 DM particles or $1.8 \times 10^5 \rm M_{\sun}$). We label this time as the time of the halo's birth, $t_{\rm birth}^h$. We define the age of the halo when it was first found to host a DCBH as 

\begin{equation}
\tau = t_{\rm DC}^h - t_{\rm birth}^h \ ,
\end{equation}
where $t_{\rm DC}^h$ is the time when a halo is found to host a newly formed DCBH.

\begin{figure}
\centerline{\includegraphics[width=0.5\textwidth,height=0.55\textwidth]{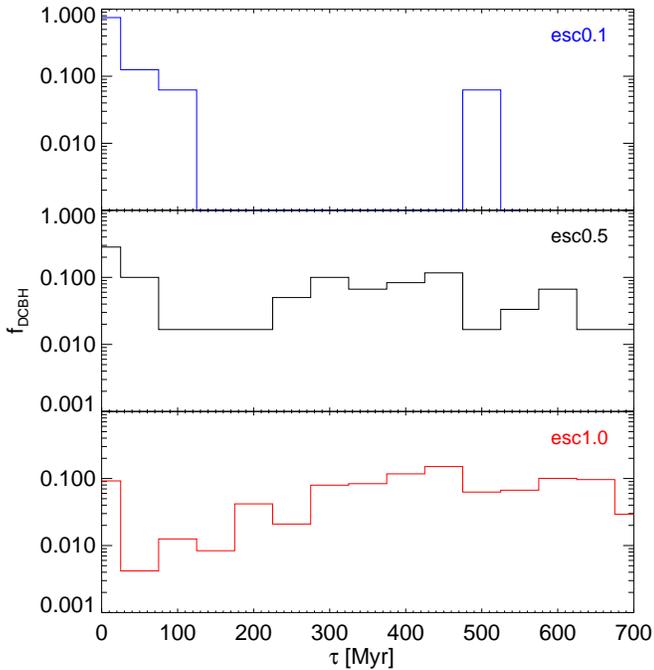}} 
\caption{Age distribution of the DCBH host haloes in our three main cases.}
\label{fig.history}
\end{figure}

The histograms of $\tau$ for our 3 main cases are plotted in in Fig. \ref{fig.history}. In the case of esc0.1 (blue), $90 \%$ of the haloes hosting a DCBH were born within 150 Myr, with the remaining haloes being 500 Myr old. However, for esc0.5 (black) and esc1.0 (red), all the DCBH host haloes are distributed over the age parameter. Part of the reason for this changing trend is that a larger $f_{\rm esc}$ implies a faster build-up of the LW flux to larger values which leads to a higher suppression of Pop III star formation over a longer time. The aforementioned suppression of Pop III star formation would also be in minihalos formed at earlier times which could later grow into massive haloes and form a DCBH if they have a close by Pop II source. Thus while in the case of esc0.5, $41 \%$ of the haloes are less than 150 Myr old, only $11 \%$ of similarly aged haloes are seen in esc1.0.

\subsection{Efficiency of DCBH formation}
\begin{figure}
\centerline{\includegraphics[width=0.5\textwidth,height=0.735\textwidth]{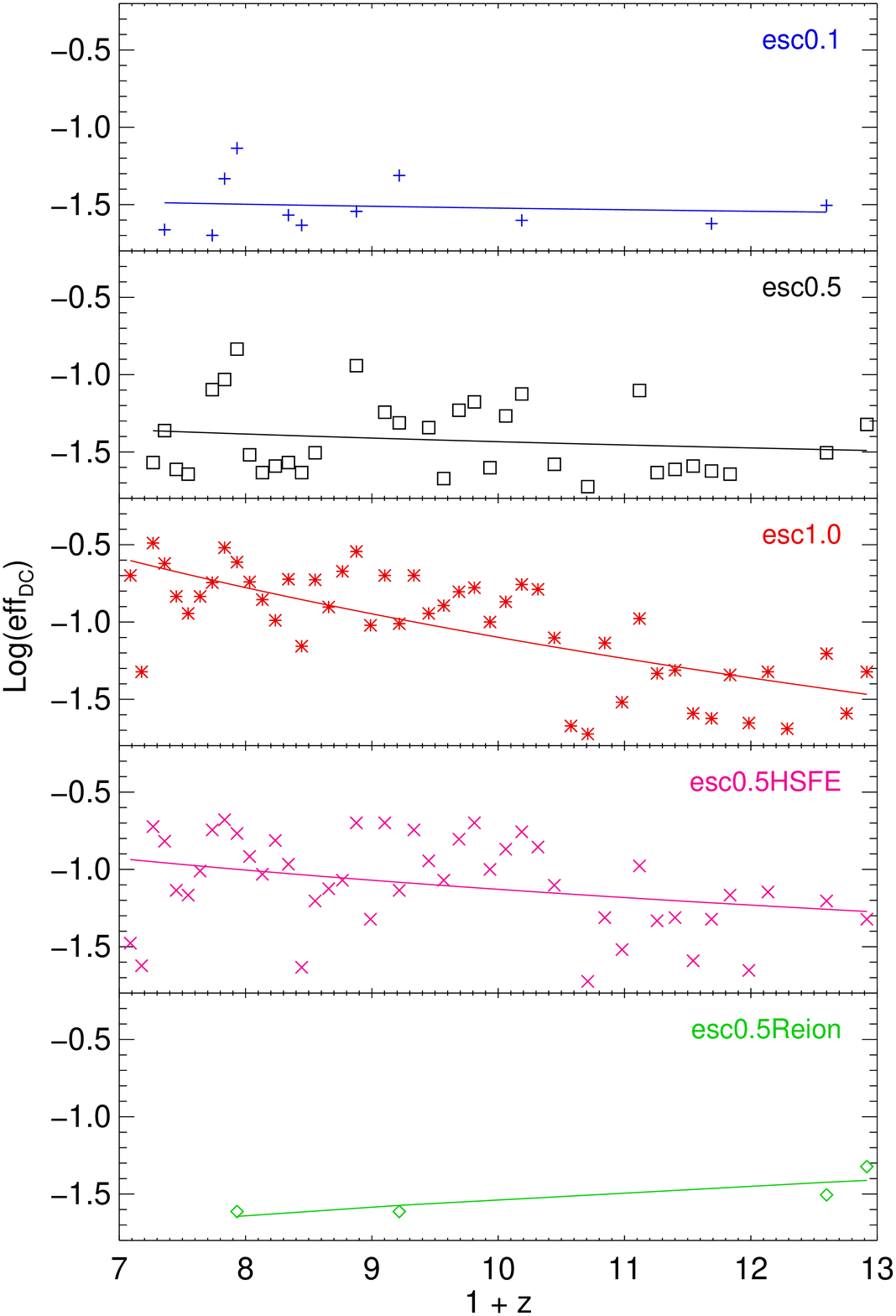} }
\caption{The efficiency of DCBH , measured as the ratio of the number of DCBHs and the total number of newly formed massive halos, as a function of redshift. The fit parameters to the lines are listed in Table \ref{tabledcbhefffit}.}
\label{fig.dcbheff}
\end{figure}

We plot the redshift evolution of the efficiency of DCBH formation, $\mbox{\textit{eff}}_{\rm DC}$ in Fig. \ref{fig.dcbheff}. The efficiency at a redshift is defined as the number of newly formed DCBHs (which by definition form in pristine massive haloes) divided by number of newly formed massive haloes at that redshift. Note that in our model a newly formed pristine halo with $T_{\rm vir} \ge 10^4 \ \rm K$ at a given redshift will immediately form either a DCBH (if $J_{\rm LW} \ge J_{\rm crit}$) or a Pop III star (if $J_{\rm LW}<J_{\rm crit}$). 
The $\mbox{\textit{eff}}_{\rm DC}$ can be expressed as a function of redshift
\begin{equation}
\mbox{\textit{eff}}_{\rm DC} = e_1 \ (1+z)^{e_2} \ ,
\label{eq.dcbheff}
\end{equation}
where the fit parameters $e_1$ and $e_2$ are listed in Table \ref{tabledcbhefffit} for all our five cases.
\begin{table}
\caption{Fit parameters to Eq.\ref{eq.dcbheff} for each case.}
\begin{center}
\begin{tabular*}{0.45\textwidth}{@{\extracolsep{\fill}}lcc}
\hline
Case & $e_1$ & $e_2$\\
\\
\hline
esc0.1 & 0.055& -0.26\\
esc0.5 & 0.12 &-0.51\\
esc1.0 & 170 &-3.32\\
esc0.5HSFE &1.4&-1.29\\
esc0.5Reion & 0.0023 & 1.1\\
\hline
\end{tabular*}
\end{center}
\label{tabledcbhefffit}
\end{table}
Since the efficiency of DCBH formation is a combination of DCBH formation rate and the formation rate of newly formed massive haloes (which is the same in all the cases), the trends in Fig. \ref{fig.dcbheff} are similar to the ones in Fig. \ref{fig.ndens}. Again, the same reasoning that applies to the DCBH formation rate, applies to the trends in the efficiency. A higher number of LW photons leads to a higher efficiency of DCBH formation. The case1.0 has the highest efficiency of DCBH formation followed by esc0.5HSFE which is due to the fact that a higher output of LW photons is seen in the former case than the latter at $z<11$ (see Fig. \ref{fig.jnetmean}). However, the efficiency in esc0.5Reion decreases at later times which is in accordance with the low overall DCBH formation rate for this case and due to the flattening in the formation rate of pristine haloes we find at later times. Note that the similar values of the fit parameters for esc0.1 and esc0.5 is due to two reasons: the formation rate densities of DCBHs in these two cases are similar and the formation rate of massive pristine haloes in all the cases is exactly the same since it is drawn from the same \textit{N}-body simulation.

For the first time, we are able to constrain the seeding mechanism of BHs at high redshifts by self consistently accounting for the physical processes that give rise to the conditions for massive BH seed formation.  Equation \ref{eq.dcbheff} encapsulates information about the number of massive metal free haloes appearing at a given epoch, their clustering with the sources (or haloes) and the rate at which these DCBHs form.

\subsection{Reionisation Feedback}
\label{sec:reion}

 We ran our fiducial case with the addition of a simple reionisation feedback prescription motivated by \cite{Dijkstra:2004p775}. During reionisation, the atomic H ionising photons ionise the gas in  massive haloes (predominantly comprised of atomic hydrogen) which results in the delayed collapse of gas. As a consequence a larger potential well is required for gas collapse due to the added gas pressure from photoheating.  In accordance with \cite{Dijkstra:2004p775}, a circular velocity threshold of $V_{c}=20 \ \rm km\, s^{-1}$ was added on top of the fiducial model to account for the reionisation feedback in haloes with $T_{\rm vir}>10^4\ \rm K$ between $6<z<11$.  In essence, this models the impact of instantaneous reionization at $z\approx 11$.
 
 The results reflect an immediate quenching of Pop III SF at $z<11$ in  massive haloes in Fig. \ref{fig:sfrI} . The mini-haloes are also unable to make Pop III stars due to the high $J_{\rm bg}$ already in place but the Pop II SFR remains almost unaffected. This is because our Pop II SF threshold already requires a halo to have $M_{\rm infall}>10^8 \ \rm M_{\sun}$ which roughly translates to a $V_c\approx 20\ \rm km\, s^{-1}$.

The most interesting outcome is the appearance of only $4$ DCBHs in our box, as compared to the 59 in our fiducial case of esc0.5, with no DCBHs seen between $8<z<11$. This accounts for the effect where even though a pristine massive halo in this redshift range might be exposed to $J_{\rm crit}$, most of the gas would be in a hot ionised state which would prevent it from collapsing and forming a DCBH. Only once the halo has a circular velocity greater than $20\rm \ \rm km\, s^{-1}$, the gas inside it can collapse and form a DCBH, which happens in our box at $z<8$. The DCBHs that form before the onset of reionisation at $z>11$ are the ones found in pristine massive haloes with no constraints on their circular velocity. This is one reason why even though the green and black lines trace each other in Fig. \ref{fig.jnetmean}, only 4 DCBHs are seen in the esc0.5reion case as compared to the 59 in the esc0.5 case. 

\begin{figure}
\leftline{\includegraphics[width=0.5\textwidth,height=0.34\textwidth]{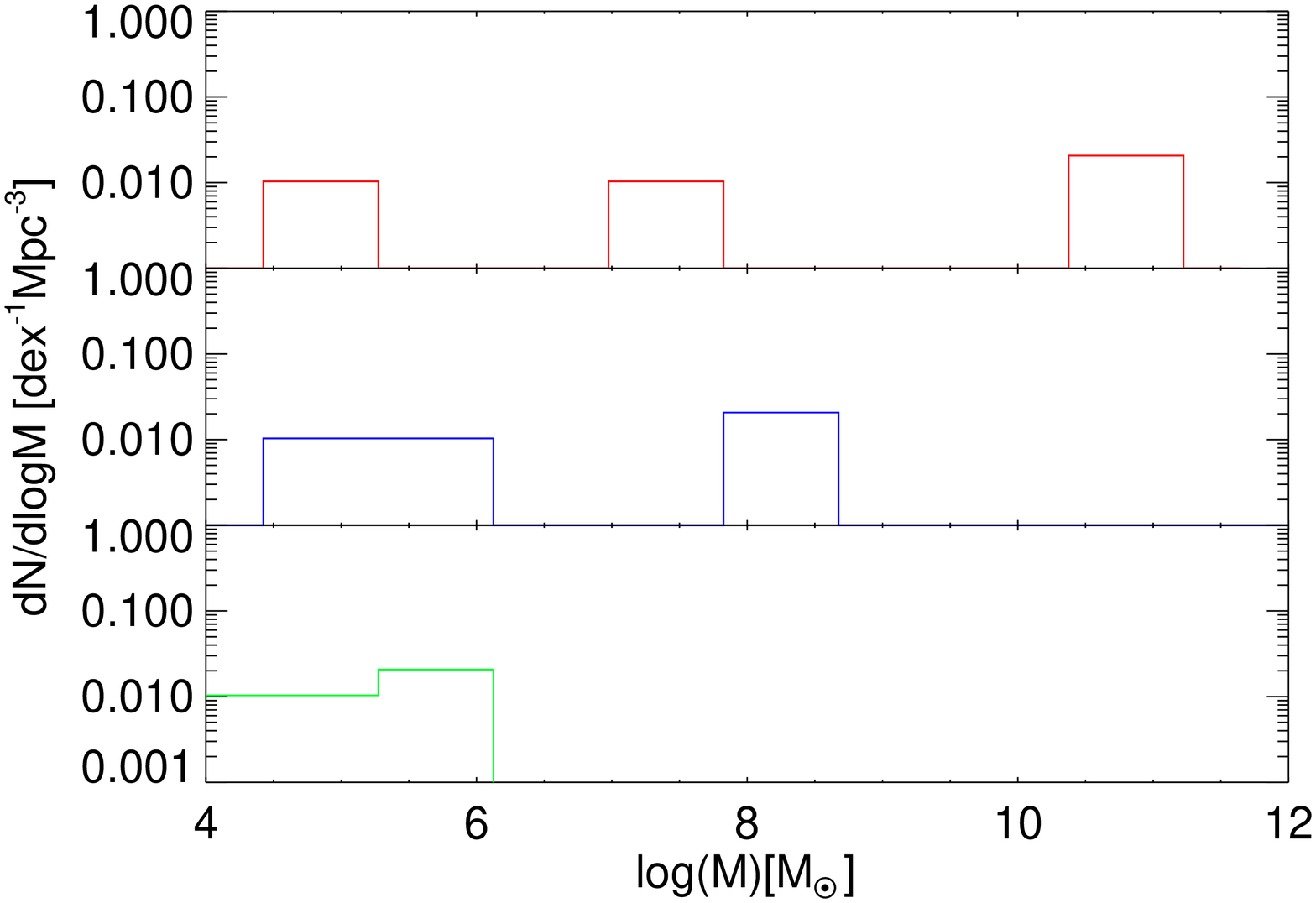}} 
\leftline{\includegraphics[width=0.5\textwidth,height=0.34\textwidth]{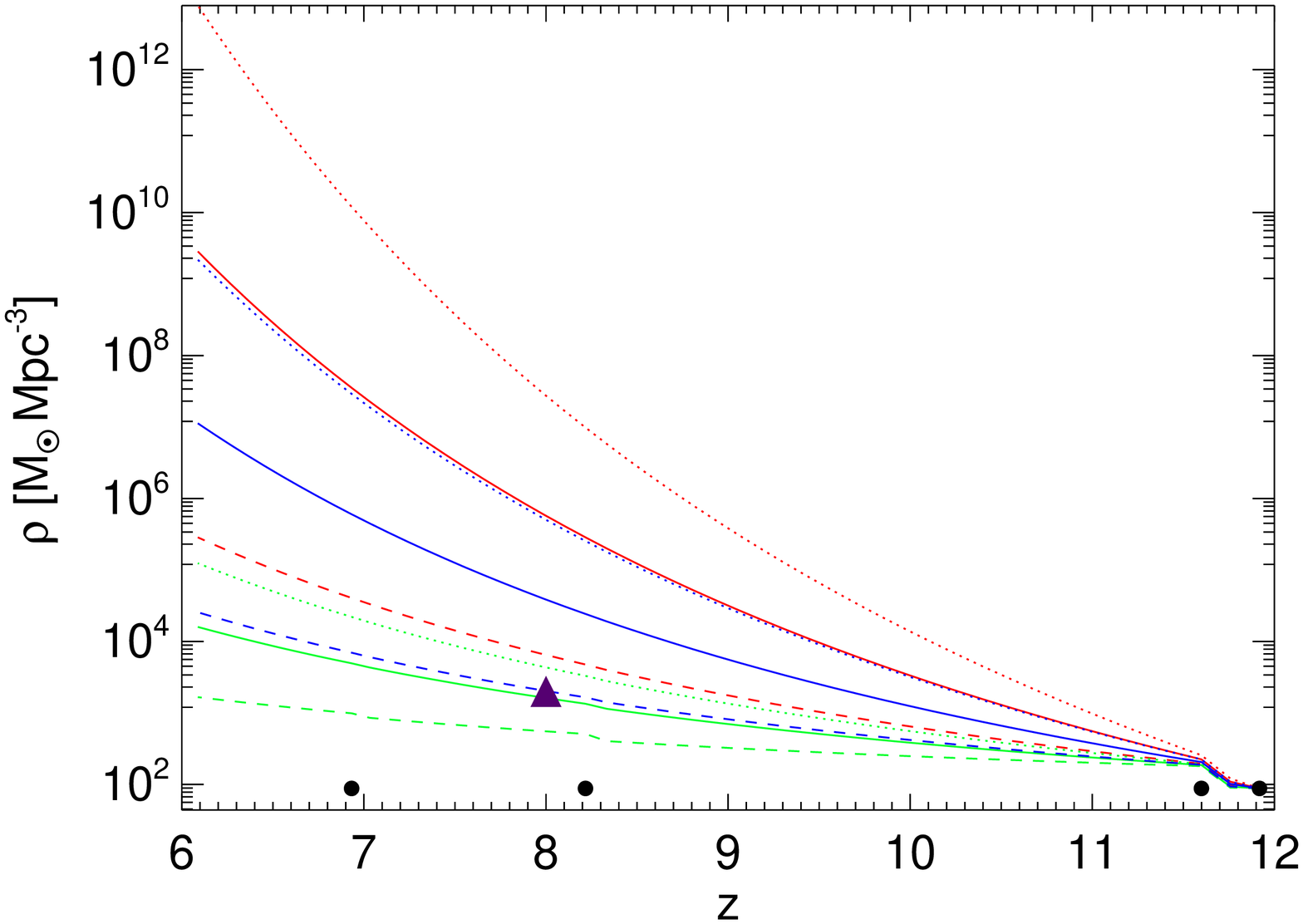}}
\caption{Same as Fig. \ref{fig.bhtwin} but for the esc0.5Reion case. The presence of only 4 DCBHs has severe consequences on the BH mass function. We are able to match the claim of T11, marked as the solid purple triangle, with both Eddington and sub-Eddington accretion modes. The zero points in the mass function arise due to the lower (factor of 10) number of DCBHs in the esc0.5reion case as compared to esc0.5.}
\label{fig.bhtwinreion}
\end{figure}

We allowed the DCBHs in the esc0.5Reion run to grow in the same way as the in esc0.5 run (described in Sec. \ref{sec:bh}). The early appearance of the 3 DCBHs at $z>11$ in this case allows them to grow into SMBHs (see Fig. \ref{fig.bhtwinreion}) by $z\sim6$ with $f_{\rm edd}=1$. 

We match the recent claim made by T11 for a population of obscured IMBHs, at $z\approx8$, by setting the $f_{\rm edd}=0.1$ and $\epsilon=0.2$ or $f_{\rm edd}=0.4$ and $\epsilon=0.1$. It is interesting to that we are able to match T11's claim with a sub-Eddington efficiency in the case where we find the least number of DCBHs.

The fact that DCBH-hosting halos are clustered, also suggests that they form in regions of the Universe that are reionised at relatively early times, as well, due to the concentration of ionising sources around them.  This suggests that the feedback from reionisation could be even stronger than what we have found assuming instantaneous reionisation at $z$ = 11.

\section{Observability of the Stellar Seeds of Direct Collapse Black Holes}
\label{sec:Observations}
We have found that a significant number of direct collapse black holes are likely to have 
formed in the early Universe.  Now we turn to the question of whether these objects are
plentiful enough for future surveys to detect them.  As discussed by e.g. \cite{Bromm:2003p22} and \cite{Begelman:2010p872}, the hot protogalactic gas is expected to first collapse to a supermassive
primordial star which subsequently accretes gas until it attains a mass of $\ga$ 10$^4$ 
M$_{\odot}$ and collapses to a black hole \cite[see also][]{Dotan:2012p876, Hosokawa:2012p886, Johnson:2012p874}.  
Here we focus on the prospects for uncovering these supermassive stellar progenitors of direct collapse black holes,
as these objects are expected to be very bright and possibly detectable by JWST \citep[e.g.][]{Gardner:2006p873}.  We shall address the question of the detectability of accreting 
direct collapse black holes in future work.

In order to estimate the likelihood that a given deep survey could find  SMSs, the precursors of DCBHs, we use the fits 
provided in Table 3 to the rate $|dN/dz|$ of SMS formation (equal to the rate of black hole formation) shown
in Fig. 7.  With this, we find that the expected number $n_{\rm SMS}$ of SMS that lie within a region of the sky, as a function 
of redshift $z$, is given by

\begin{eqnarray}
\frac{dn_{\rm SMS}}{dz} & = & \frac{dV}{dz} \left|\frac{dz}{dt}\right|  \left|\frac{dN}{dz}\right|  t_{\rm life} \nonumber \\
   & \simeq &  100 \, {\rm deg}^{-2} \, \left(\frac{1+z}{10} \right) \nonumber \\
& \times & \left(\frac{|\frac{dN}{dz}|}{10^{-3} \, {\rm Mpc^{-3}}} \right) \left(\frac{t_{\rm life}}{10^6 \, {\rm yr}} \right)\mbox{\ ,}
\end{eqnarray}
where $dV/dz$ is the comoving volume element per unit redshift, $|dt/dz|$ is the rate of change 
of the Hubble time with redshift, and $t_{\rm life}$ is the lifetime of a SMS.  To obtain the second equation above
we have neglected the effect of dark energy on the rate of Hubble expansion, which is a reasonable
assumption at the high redshifts ($z\ga 6$) we are considering here; otherwise, we have adopted 
the same cosmological parameters as described in Section 2.1.  Finally, note that the longer the 
stellar lifetime $t_{\rm life}$, the more objects will be visible within a given redshift interval.

\begin{figure}
\centerline{\includegraphics[width=0.5\textwidth]{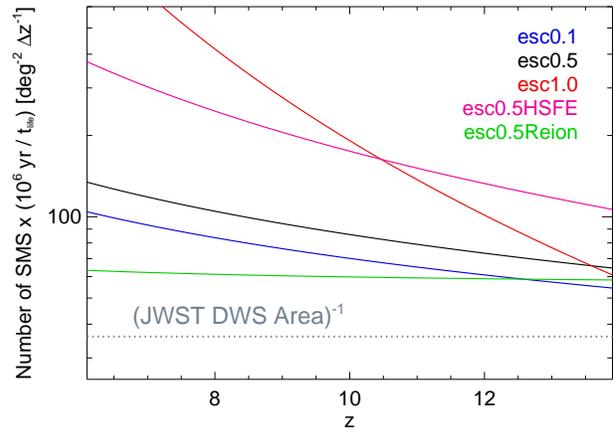}} 
\caption{The number of supermassive stellar progenitors of direct collapse black holes, as observed on the sky per square degree per redshift interval ($\Delta z$), as a function of redshift $z$.  The number of supermassive stars is given by the fits shown in Fig. 7 and in Table 3, for each of the five cases shown.  The gray dotted line shows the number of supermassive stars that must be present for at least one per redshift interval ($\Delta$$z$ = 1) to appear in the field of view of the Deep-Wide Survey planned for the JWST.  For all cases the survey should be large enough for at least a few to of the order of 10  supermassive stars to lie within the field of view.  Strong continuum and H$\alpha$ and He~{\sc ii} $\lambda$1640 emission lines may be detected from these objects.}
\label{fig.jwst}
\end{figure}

Fig. \ref {fig.jwst} shows the number of SMS per square degree per redshift interval that we find for 
each of the five cases shown in Fig. 7, normalized to $t_{\rm life}$ = 10$^6$ yr, a typical value 
expected for a rapidly accreting SMS \cite[see][]{Begelman:2010p872,  Johnson:2012p874}.  
Also shown is the minimum number of SMS that would yield an average of one SMS 
per redshift interval ($\Delta z = 1$) within the area of sky covered by the Deep-Wide Survey (DWS) 
planned for the JWST \cite[e.g.][]{Gardner:2006p873}.  Clearly, the prospects of detection are good, 
as in each of the cases we find that at least a few SMS should lie within the survey area $\sim 100$ arcmin$\times$arcmin.

We note that rapidly accreting SMS are expected to have distinct observational signatures
which could be detected by the JWST, as discussed by  \cite{Johnson:2012p874}.  In particular, 
these objects may emit strong continuum
radiation below the Lyman limit, and they are likely to exhibit both strong H$\alpha$ and He~{\sc ii} $\lambda$1640
recombination line emission.  An important difference between these objects and others with strong recombination
line emission is that they may also be very weak Ly$\alpha$ emitters, due to the trapping of Ly$\alpha$ photons 
in the optically thick accretion flows feeding their growth.  The detection of objects exhibiting these
 observational signatures would provide important constraints on both the nature and abundance of the stellar seeds
of direct collapse black holes.

 \section {Summary and Discsussion}
 \label{sec:Discussion}
In this paper, we present the results from a $N$-body, DM only simulation of a 3.4 Mpc$h^{-1}$ co-moving box from cosmological initial conditions. On top of this simulation, we developed a SAM which takes into account the self consistent global build up and local variation of the LW radiation field due to the Pop III and Pop II stellar sources. The merging histories of haloes are also tracked in order to account for metal pollution form previous episodes of star formation. This allowed us to identify the possible sites of DCBH and to investigate their environment. Though our simulation is not large enough to probe a wide range of environments, we show that even in such volumes, BH seeding by DCBHs could be a common phenomenon. Our study in this respect motivates the seeding of present-day SMBHs via the formation of DCBHs. The key findings of our work are summarised below.

\begin{enumerate}

\item DCBH formation sets in with the onset of Pop II star formation as the Pop II stars can easily produce $J_{\rm LW}\ge J_{\rm crit}^{\rm II}$. On the other hand, LW radiation from Pop III stars is not able to exceed $J^{\rm III}_{\rm crit}$. 
\\
\item We find the first DCBH at $z\approx12$, however the total number of such objects depends on the LW photon output of a given model.
\\
\item In each of our cases, all the haloes that host DCBHs have close by LW sources within $\sim 10$~kpc. 
\\
\item We also find that the haloes that host DCBH at $z > 10$ are more clustered with external LW sources than the haloes that do not host DCBH.
Also, the DCBHs that appear later at $z<10$ are less clustered than the DCBHs that appear at $z>10$ and may exist at the centres of galaxies of various morphological types at $z=0$.
\\
\item In our model including reionisation, we are able to match recent claims made by T11 about the population of obscured IMBHs, by assuming both Eddington ($f_{\rm edd} = 0.1$, $\epsilon = 0.2$) and sub-Eddington ($f_{\rm edd} = 0.4$, $\epsilon = 0.1$) accretion modes for the DCBHs.
\\
\item We find that for all our cases, the JWST should be able to detect at least a few of the supermassive stellar precursors of these DCBHs over a wide range of redshifts ($z>6$).

\end{enumerate}

Our results are subject to limitations due to the modelling approach we chose. The halo threshold mass assumed in our work for Pop II star formation sets the clock for DCBH formation. In our current work, we have set the mass threshold for Pop II star formation to $10^8\ \rm M_{\sun}$ following the work of \cite{Maio:2011p104}, however setting it to a lower value would allow for the Pop II stars to form earlier in the box. This would lead to an earlier epoch of DCBH formation but it is difficult to predict their abundance at later times. Note that we also set the mass threshold for structure formation (Pop III, Pop II or DCBH) to $V_{\rm c}=20\ \rm km\, s^{-1}$ between $6<z<11$, which only further quenches the DCBH formation rate.

Note that our simulated volume is smaller then typical volumes probed by current observations and does not include sources as luminous as the ones detected in the surveys. Thus our predicted SFRD should be somewhat lower than the observed ones. Based on the observational constraints, as shown in Fig. \ref{fig.smitmf} and \ref{fig:sfrI}, we estimate that our computed value of $J_{\rm bg}$ could be lower by a factor of $\simeq 2$ at a given redshift. A higher level of LW background would make it relatively easy to reach $J_{\rm crit}$ and would also imply the quenching of Pop III star formation in a larger number of minihaloes. Whether this would also lead to a higher number of DCBHs is non-trivial to predict.

In principle, SN explosions from neighbouring stellar populations can enrich a pristine halo early on in its lifetime \citep{Maio:2011p104}. This could also reduce the number of DCBHs we find in our study, if the metal enrichment is high enough to alter the cooling properties of the gas \citep{Omukai:2008p113}. However, it is very likely that the metals carried in the SN wind may not be mixed into the dense gas at the centre of the halo where DCBH formation occurs \citep[e.g.][]{Cen:2008p841} within the timescale of DCBH formation which is $\approx 2-3 \ \rm Myr$.

The gas within the haloes identified as DC candidates would still need to collapse without fragmentation into a central massive object. The study by LN06 explores a mechanism where a Toomre-stable gaseous disc in a pristine low spin halo can effectively redistribute its angular momentum, thereby preventing fragmentation and eventually forming a DCBH. The aim of our next study is to self consistently explore the mechanism suggested by LN06, on top of our existing framework, which should in principle greatly reduce the number of DCBH host haloes since low spin haloes at such high redshifts are quite rare \citep[eg.][]{Davis:2010p139}.

The accreting discs of both Pop III remnant BHs and DCBHs could also emit LW photons \citep[e.g.][]{Pelupessy:2007p1050, Alvarez:2009p778, Jeon:2011p1083}, where the emission would depend on both the BH mass and accretion rate \citep{Greif:2008p377}. However, a recent study \citep{Johnson:2011p704} has shown that due to the low accretion rate of these BHs, their contribution to the LW specific intensity can be quite low outside the halo at $\sim 1\ \rm kpc$, i.e. the typical virial radius of BH host haloes at high redshifts. Due to the uncertainty in the emission characteristics of BH accretion disc, we focus on the stellar components to account for the LW radiation in our model.  Also, the X-ray feedback from accretion discs could heat the gas in surrounding haloes, thereby preventing them from collapsing and making stars \citep[e.g.][]{Mirabel:2011p1115, Tanaka:2012p1114}, hindering the formation of DCBHs in the neighbouring pristine massive haloes. We plan to explore the impact of accreting BHs on the formation of DCBHs in a future study.

Since the plausibility of direct collapse in a pristine halo depends on the number of LW photons reaching it, we find a clear degeneracy in the various cases that span the $(f_{\rm esc},\alpha)$ parameter space. The degeneracy in the number of LW photons produced in our cases could be broken by comparison of the BH mass function or the mass density of BHs that we find with those inferred for BHs from observations at $\rm z > 6$. Another possibility is via the detection of SMSs in the planned surveys of the JWST which could also shed light on the plausibility of this scenario as a SMS is believed to be the precursor of a DCBH \cite[e.g.][]{Begelman:2010p872}.
  
Our results shed new light on the long-standing argument that only very close star-halo pairs could give rise to DCBHs and that a characteristic length of $\sim 10$~kpc is the maximum distance within which a halo must see a LW source in order to have direct collapse of gas (D08). We find that although a source must exist within 10 kpc, it is not necessary that a single source produces all of the LW radiation which accounts for $J_{\rm crit}$; there is a contribution from the cosmological background LW radiation field, as well as from a number of local sources producing $J_{\rm crit}$. However, in all the cases, it is only the Pop II star clusters that produce all (or most) of the $J_{\rm crit}$. Pop III stars alone never produce enough LW photons to achieve $J_{\rm crit}$.
 
 It is  interesting to note that even in the worst-case scenario for the formation of DCBHs i.e. the model including reionisation feedback (esc0.5Reion), we still find a few DCBHs, which hints towards the high plausibility of the DCBH scenario. While photoionisation strongly inhibits the formation of Pop III stars in smaller pristine haloes, it still allows for Pop II star formation in massive enough enriched haloes, which produce the necessary background. 
 
Allowing the DCBHs to grow via different modes of Eddington accretion gives rise to a range of possibilities for the BH mass function, and can readily account for the presence of supermassive black holes by $z=7$. The expected number of SMBH is a few per co-moving $\rm Gpc^3$, in accordance with the inferred number of quasars at $z>6$ \citep{Fan:2003p40,Fan:2006p149,Mortlock:2011p447}. We over predict the number of such SMBHs but argue that our work is an upper limit for the existence of such objects. However, the study by T11 \citep[see also][]{Willott:2011p448, Fiore:2012p834} suggests the possibility of a large number of intermediate mass black holes at $z>6$. They infer (via extrapolation) the presence of an obscured population of intermediate mass BHs by looking at the stacked X-ray luminosity signals of high redshift galaxies. We are able to match their claim at $z\sim8$ in our reionisation model, assuming both Eddington and sub-Eddington accretion modes for the DCBHs. Independent of the claim made by T11, on the basis of our model we argue that a population of BHs must be present at $\rm z>6$ due to the sheer number of DCBH host haloes that we find. 

A precise seeding mechanism of BHs at early redshifts in cosmological simulations is important in order to explain the AGN luminosity functions, growth of massive BHs and the evolution and properties of galaxies at lower redshifts. The environment of these BHs would play an important role in determining their evolution and the Eq. \ref{eq.dcbheff} is a first step towards constraining the environments and masses of seed BHs. Using a semi-analytical model, which takes into account the halo histories and the spatial variation of the LW flux, we were able to parameterise the fraction of  newly formed haloes with $T_{\rm vir} >10^4\ \rm K$ that are able to host DCBH as a function of redshift. The equation is an outcome of our model where we are able to resolve haloes with masses in the range $10^{6-7}\ \rm M_{\sun}$ and could thus serve as a sub-grid model for the seeding of BHs in large scale cosmological simulations, which we will pursue in a future study. 

\section*{Acknowledgments}

The authors gratefully acknowledge the anonymous referee for her$/$his useful comments. The authors would like to thank Volker Springel for allowing them to use \textsc{gadget} and \textsc{subfind}. The authors gratefully acknowledge Priyamvada Natarajan for her very valuable comments during the final stages of the paper. BA acknowledges the useful discussions with Stefanie Phleps during the early stages of the work. BA would like to thank Umberto Maio, Fabrice Durier and the TMoX group at the MPE for the constructive criticism. The authors are grateful to the support staff of the SFC supercomputer cluster at the Rechenzentrum Garching of the Max Planck Society, on which the simulations presented here were carried out. JLJ gratefully acknowledges the support of the U.S. Department of Energy through the LANL/LDRD Program for this work. SK acknowledges support from the the Royal Society Joint Projects Grant JP0869822. This work acknowledges support from the DFG Cluster of Excellence ÔOrigin and Structure of the UniverseÕ. CD acknowledges the support from the Marie Curie Reintegration Grant FP7-RG-256573.
 \bibliographystyle{mn2e}


\appendix
\section[]{Details of Methodology}

\subsection {Mass Function at $z=6$}
\label{sec:MF}
\begin{figure}
\centerline{\includegraphics[width=0.5\textwidth,height=0.34\textwidth]{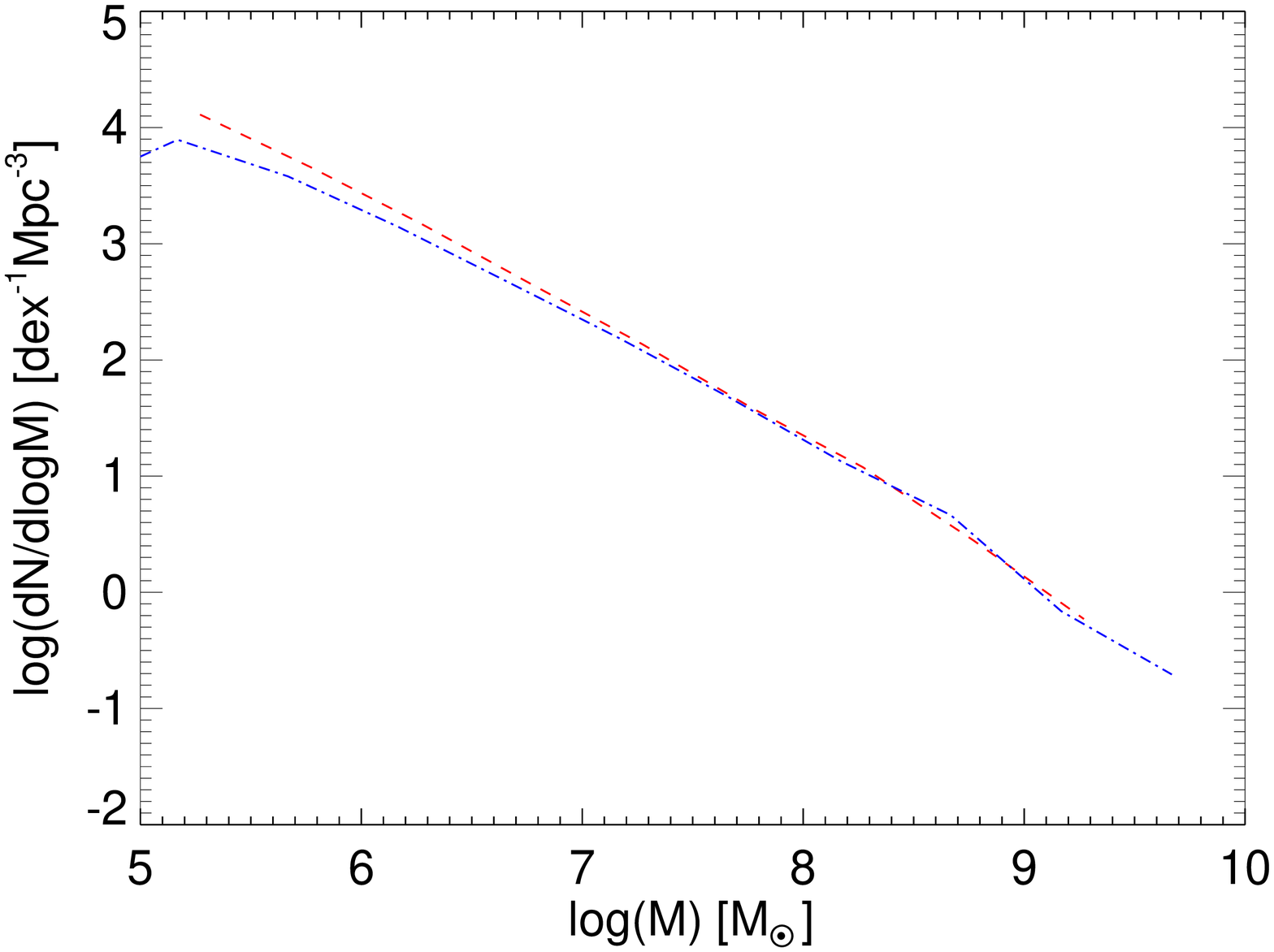}} 
\caption{The mass function of haloes at $z=6$. The red dashed line depicts the mass function of subhaloes and blue dashed-dotted line represents the FoF mass function in our work.}
\label{fig.MF}
\end{figure}
We plot the mass function of the haloes in our work at $z=6$ in Fig. \ref{fig.MF}. The red and blue lines depict the subfind and FoF halo mass function in our work respectively. It is clear from the plot that we probe the low mass end of the mass function at $z=6$.
\subsection {Selection of LW sources}
\label{sec:selection}
The stellar populations (both Pop III and Pop II) are the primary source of LW radiation at early epochs. However, only certain sources can contribute to a LW radiation background at any given snapshot. Two important processes govern the fate of these LW photons; first, they might get cosmologically redshifted out of the LW band while simultaneously, the photons from the bluer end of the spectrum enter the LW range and second, these photons can get absorbed by the neutral hydrogen present in the early Universe. \cite{Haiman:2000p87} looked at the absorption of these photons by the neutral hydrogen present in the Universe. They concluded that the $\Delta z_{\rm LW}$ over which a LW photon can exist is quite small since it gets readily absorbed by atomic hydrogen present in the un-reionised Universe. The LW band range lies very close and even overlaps with transitions that occur in atomic hydrogen, hence the mean free path for a LW photon is smaller than the distance it can travel before it gets cosmologically redshifted out of the band \cite[][Fig. 16]{Haiman:2000p87}. 
Using the relation
\begin{equation}
\label{eq.redshift photons}
\frac{1+z_{\rm max}}{1+z_{\rm obs}}=\frac{\nu_{\rm i}}{\nu_{\rm obs}} \ ,
\end{equation} 
one can easily compute the maximum redshift ($z_{\rm max} > z_{\rm obs}$) at which a photon emitted at frequency $\nu_{\rm i}$ can contribute to the LW band, at a given observation redshift ($z_{obs}$) for a given observation frequency ($\nu_{obs}$). The upper limit on the lookback redshift ($z_{\rm lb}$) or lookback time ($t_{\rm lb}$) can be obtained by setting $\nu_{\rm i}=12.1 \ \rm eV$ (owing to the Lyman-$\beta$ line) and $\nu_{\rm obs}=11.2 \ \rm eV$ for any given redshift. Hence while calculating the mean LW background for a given redshift, we only count the stars whose photons originate after $t_{\rm lb}$. To do this, we define two important parameters for each star/stellar population  in our study; the time of formation, which refers to the age of the Universe when the star was formed denoted by $t_{\rm form}$ and the age of the Universe when the star died denoted by 
\begin{equation}
t_{\rm contrib}=t_{\rm form} +  t_{\rm life} \ .
\end{equation}
The lifetime $t_{\rm life}$ of the star depends on the mass of the star and is computed using the fits mentioned in Table \ref{table.func fits} for Pop III stars (typical $t_{\rm life}$ of a $100 \ \rm M_{\sun}$ star $\approx 2.79 \ \rm Myr$) and \cite{Padovani:1993p67} for Pop II star cluster (typical $t_{\rm life}$ for a Pop II cluster weighted by IMF used $\approx 10 \ \rm Gyr$). Hence, the selection criteria for sources contributing to the $J_{\rm bg}$ becomes
\begin{equation}
\label{eq.meanselect}
t_{\rm contrib} \geq t_{\rm lb} \mbox{\ .}
\label{eq.0select}
\end{equation}
In addition to Eq.\ref{eq.0select}, the selection criteria for stars that can contribute locally to the LW radiation level also needs to be considered. In order to do this, we use a similar approach to KA09,and analyse the past light cone of a halo and compare it the world lines of the sources. At a given timestep $t_i$, we check if LW photons from a source can contribute to the $\rm J_{local}$ in a halo by comparing the physical distance between the source and the halo's position, with the time required for the radiation to travel between the birth of the source and $t_i$, and the death of the source and $t_i$. In case of emission from Pop II stellar clusters, we also calculate their age in order to determine \textit{when} the photons were actually emitted (Fig. \ref{fig.ST99}).

\begin{figure}
\leftline{\includegraphics[width=0.34\textwidth,height=0.5\textwidth,angle=270]{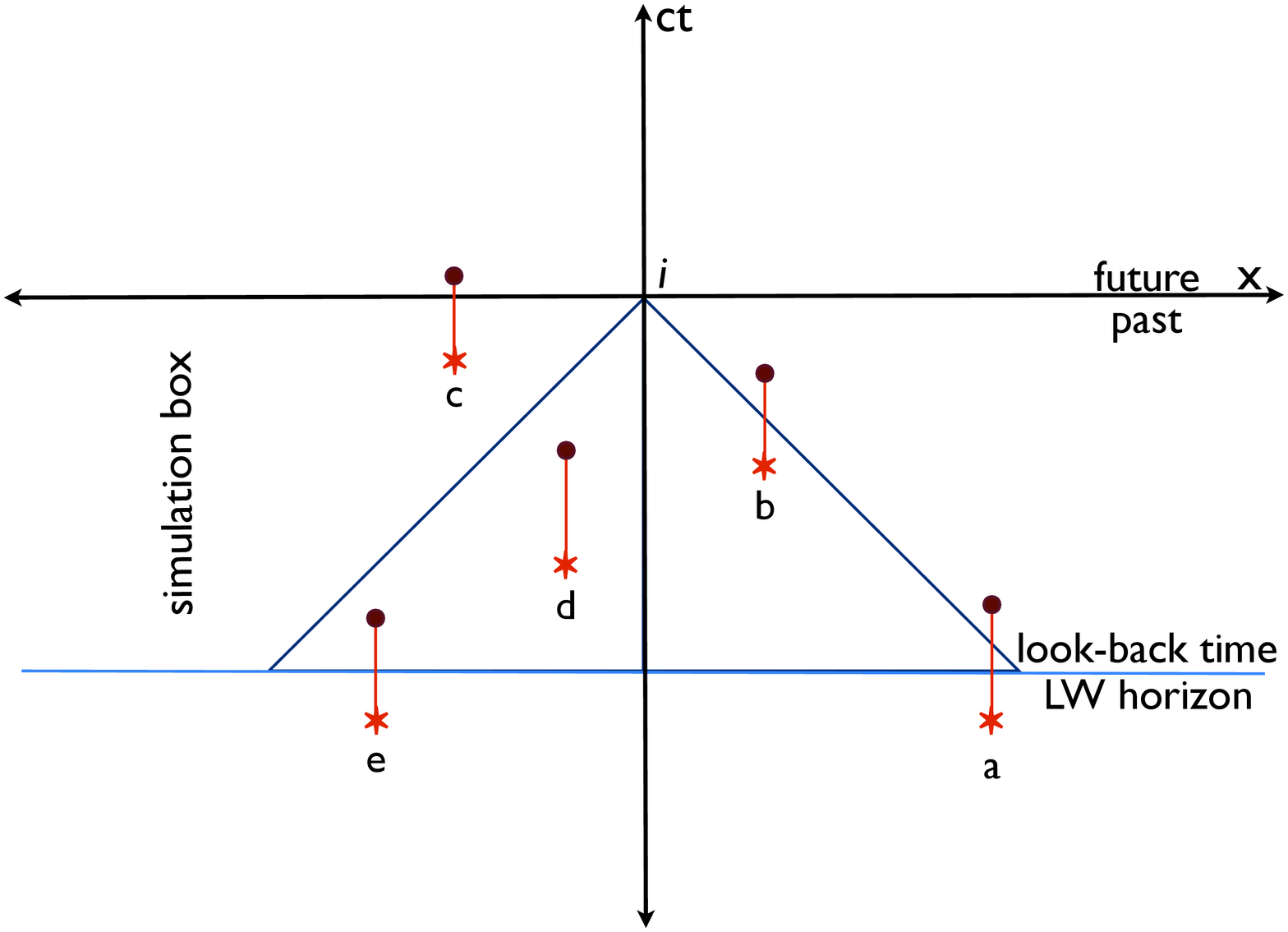}}
\caption{Lightcone diagram for the selection criteria of LW sources in our work. Red stars indicate the $t_{\rm form}$ and maroon filled circles represent the $t_{\rm contrib}$ for a stellar source. The halo for which the LW intensity is to be calculated is placed at \textit{i} whose past lightcone is marked in dark blue. Although all the sources \textit {a,b,c,d,e} satisfy the Eq. \ref{eq.0select}  and will contribute to $J_{\rm bg}$, only the LW photons from stellar sources \textit{a} and \textit{b} can make it to the halo following Eq. \ref{eq.1select} and \ref{eq.2select} and will contribute to $J_{\rm local}$.}
\label{fig.LWsel}
\end{figure}

We describe our prescription for selecting stars that are considered to contribute spatially to the LW intensity in a halo at time $t_i$ by writing the conditions
\begin{eqnarray}
\label{eq.1select}
d_{\rm s-h}      & \leq & D_{\rm lt,if} \ ,\\ 
\label{eq.2select}
d_{\rm s-h}      & \geq & D_{\rm lt,ic} \ ,
\end{eqnarray}
where $d_{\rm s-h}$ is the physical source-halo distance, $t_{i}$ is the age of the Universe at snapshot \textit{i}, $D_{\rm lt,if}$ and $D_{\rm lt,ic}$ represent the physical distance light can travel between  $t_i$ and $t_{\rm form}$ and between $t_i$ and $t_{\rm contrib}$ respectively (see Fig. \ref{fig.LWsel} for more details).
Hence if a source/star satisfies Eqs. \ref{eq.meanselect}, \ref{eq.1select} and \ref{eq.2select} then it contributes locally to the LW radiation level and the selection criteria for the sources that can contribute to $J_{\rm bg}$ is given by Eq.\ref{eq.meanselect}.

\begin{figure}
\leftline{\includegraphics[width=0.34\textwidth,height=0.5\textwidth,angle=270]{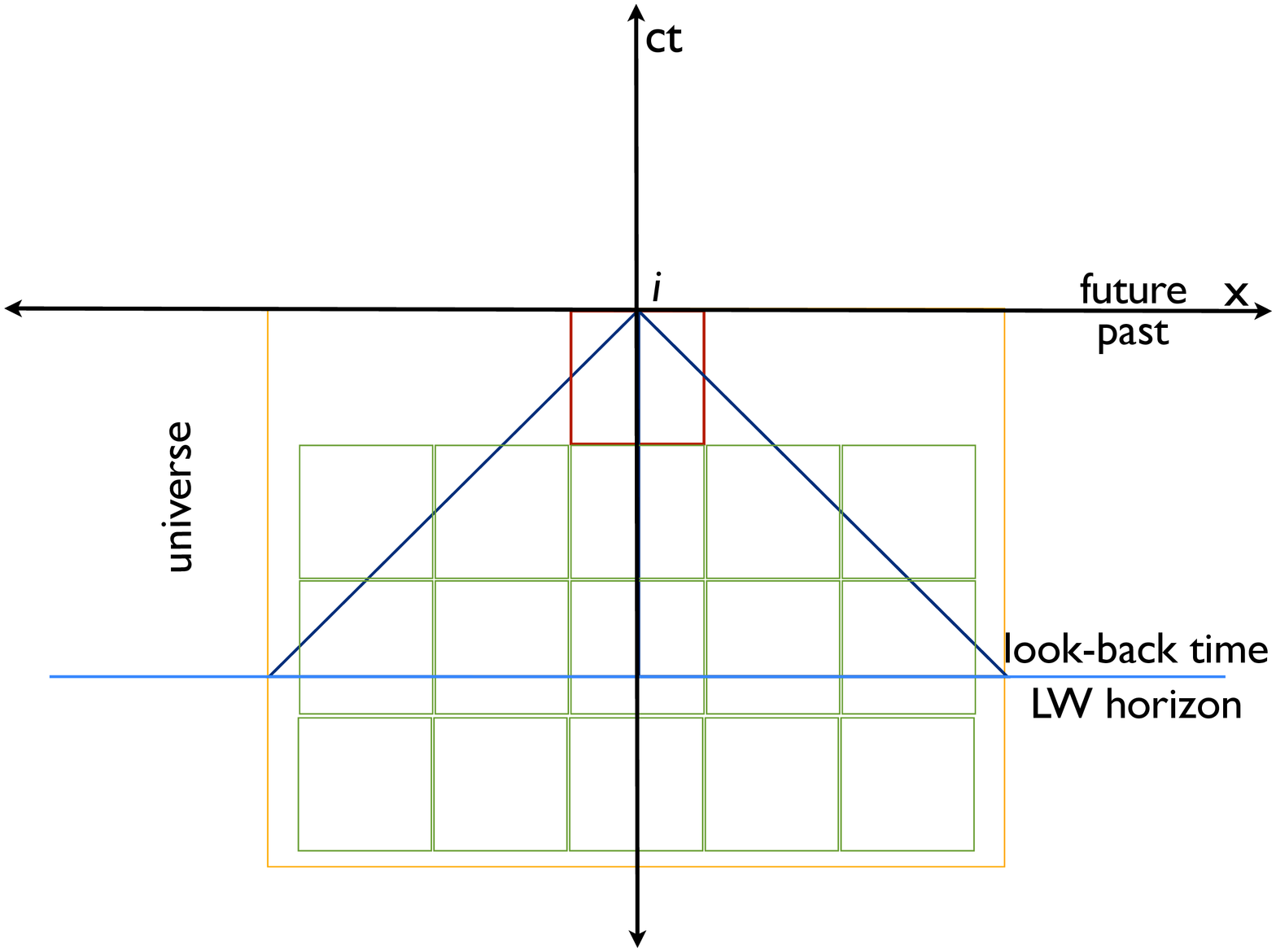}}
\caption{A halo is (periodically) placed at the centre of one of the sides of the simulation box (red) at epoch \textit{i}. The halo's past lightcone is denoted by the dark blue lines. The light blue line marks the lookback time computed using \ref{eq.redshift photons}. The actual background \Jlw would come from the entire Universe (orange box) which can in turn be imagined as a conglomerate of smaller simulation boxes (green). The background \Jlw is computed using the red box but it is assumed that the same mean \Jlw would exist throughout the Universe (orange box).}
\label{fig.Universe}
\end{figure}

It is important to note that every halo early on in the Universe is expected to be exposed to a minimum level of \Jlw given by Eq. \ref{eq.JLWIII} and \ref{eq.JLWII} which is an approximation to the mean-background level of radiation that is believed to be present \textit{everywhere} in the Universe. Hence, in Eqs. \ref{eq.JLWIII} and \ref{eq.JLWII} we assume that the SFR density in a $\sim$ few Mpc-side box (and hence the comoving density of stars) would be the same everywhere in the Universe (see Fig. \ref{fig.Universe}).  \\
\label{lastpage}

\end{document}